\documentclass[preprint]{aastex62}
\usepackage{graphicx}
\usepackage{amsmath}
\usepackage{tabularx}
\usepackage[caption=false]{subfig}
\usepackage{rotating}
\usepackage{verbatim}
\usepackage{needspace}

\newcommand\frameworkitem[1]{\paragraph{{#1}.}}

\def\Msolar{\ifmmode {\rm M_{\odot}}\else $\rm M_{\odot}$\fi}
\def\Mearth{\ifmmode {\rm M_{\oplus}}\else $\rm M_{\oplus}$\fi}
\def\Rearth{\ifmmode {\rm R_{\oplus}}\else $\rm R_{\oplus}$\fi}

\def\techo{\tau_\text{echo}}
\def\thalf{\tau_{1/2}}
\def\deltat{\Delta t}
\def\rmsnoi{\cal{N}_\text{rms}}
\def\intflarepeak{{\cal{F}}_p} % counts in bin for an impulsive flare.
\def\nflares{N_\text{f}}
\def\signoise{s_{min}}

\begin{document}

\title{A Framework for Planet Detection with Faint Light-curve Echoes}

\author{Chris Mann}
\affil{Nanohmics, Inc., 6201 E Oltorf St., Ste 400, Austin, TX 78741}
\email{cmann@nanohmics.com}

\author{Christopher A. Tellesbo}
\affil{Department of Physics \& Astronomy, University of Utah, 
\\ 115 S 1400 E, Rm 201, Salt Lake City, UT 84112}
\email{chris.tellesbo@utah.edu}

\author{Benjamin C. Bromley}
\affil{Department of Physics \& Astronomy, University of Utah, 
\\ 115 S 1400 E, Rm 201, Salt Lake City, UT 84112}
\email{bromley@physics.utah.edu}

\author{Scott J. Kenyon}
\affil{Smithsonian Astrophysical Observatory,
\\ 60 Garden St., Cambridge, MA 02138}
\email{skenyon@cfa.harvard.edu}

\begin{abstract}

A stellar flare can brighten a planet in orbit around its host star, producing a light curve with a faint echo. 
This echo, and others from subsequent flares, can lead to the planet's discovery, revealing its orbital configuration and physical characteristics.
A challenge is that an echo is faint relative to the flare and measurement noise. 
Here we use a method, based on autocorrelation function estimation, to extract faint planetary echoes from stellar flare light curves. 
A key component of our approach is that we compensate for planetary motion; measures of echo strength are then co-added into a strong signal. 
Using simple flare models in simulations, we explore the feasibility of this method with current technology for detecting planets around nearby M dwarfs.  
We also illustrate how our method can tightly constrain a planet's orbital elements and the mass of its host star.  
This technique is most sensitive to giant planets within 0.1~au of active flare stars 
and offers new opportunities for planet discovery in orientations and configurations that are inaccessible with other planet search methods.

\end{abstract}


\needspace{6em}
\section{Introduction} 
\label{intro}

Like a radar chirp or a sonar ping, a pulse of light from an
astrophysical object allows us to glimpse the surrounding darkness. By
tracking the light curve of a time-varying astrophysical source, we can
extract the echo of such a pulse, looking for scattered light from
things otherwise hidden \citep[e.g.,][]{rest2012}.  Examples include
the discovery of rings and other features following supernova
explosions \citep[e.g.,][]{crotts1989}, and reverberation mapping of
protoplanetary accretion disks around stars and gas in the vicinity of
supermassive black holes \citep[e.g.,][]{horne1991, peterson2004,
  huan2016}.  Light echoes of stellar flares may potentially reveal
the presence of planetary disks of dust/debris \citep{gaidos1994,
  sugerman2003} and planets 
\citep{argyle1974, matloff1976, bromley1992, clark2009, mann2017,
  sparks2018}. Recently, Sparks et al evaluated additional discriminants of the scattered light, including fluorescence signatures and Doppler effects, thus bolstering the case for detecting echoes around M dwarfs.\citep{sparks2018}
Echoes from multiple flare events from the same star change with the orbit of a planet and its physical properties, 
thus providing a new avenue for extracting orbital parameters\citep{argyle1974, mann2017}.

Detection of planetary echoes is challenging.  However, if it can be 
accomplished, it nicely complements other planet detection methods. 
Transit surveys, although limited to
edge-on systems, can provide planetary and stellar radii and 
estimates of orbital elements \citep[e.g.,][]{borucki2010,
  mullally2015}. Radial velocity studies \citep[e.g.,][]{mayor2011}
also constrain planetary masses; when combined with transits, they 
set a gold standard for planet confirmation.  Other methods,
including microlensing \citep[see][]{sumi2011} and direct detections
\citep[e.g.,][]{marois2008, janson2013}, are sensitive to planets on 
distant orbits.  

\begin{figure}
\centerline{%
\includegraphics[width=4.0in]{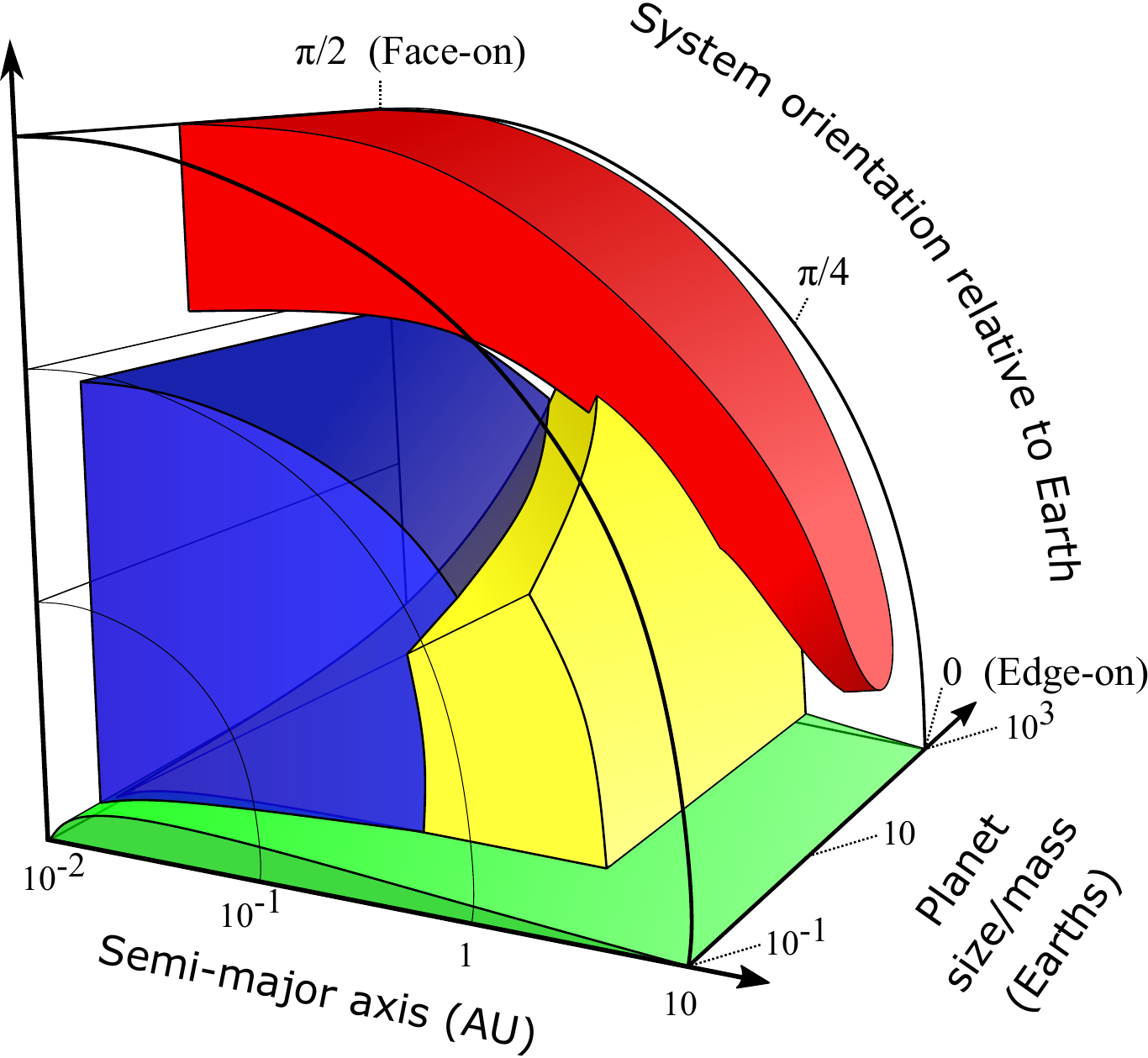}}
\caption{\label{fig:wedge}Detection regimes for each leading exoplanet
  detection method, plotted in a cylindrical coordinate system with the semi-major axis as the radial coordinate, system orientation as the azimuthal coordinate, and planet size/mass as the depth coordinate (adapted from \citep{mann2017}). 
	Transit is shown in green, radial velocity in yellow, direct imaging in red, and echo detection in blue. 
	Echoes allow probing into inner solar systems viewed from any angle, which were previously very difficult to detect.}
\end{figure}

Planetary echoes provide a direct detection method for close-in planets that 
works irrespective of a viewer's orientation, made possible only by the
time-variability of their host.  Figure~\ref{fig:wedge} illustrates
the sensitivity of these detection methods to planet size, orbital
distance, and orientation relative to an observer.

A key to using planetary echoes for new detections, or as follow-up to
known planetary systems, is repeat events. Like transits and radial
velocity measurements, sampling a planet at various points along its
orbit allows for reconstruction of orbital elements. In echo detection, 
differences in the arrival time of echoes can provide
tight constraints on orbital elements and viewing
angles. Because sets of echo delay times depend on physical distance and
orbital period, the mass of the host star can also be accurately estimated. 
That echoes are photons reflected from the planet itself means that their 
strength depends on planet radius and albedo (atmospheric and surface 
composition) and 
orbital phase. Echo sets can arise from any viewing angle, but echo
strengths fall off quickly with distance. Thus the method is most
sensitive to giant planets inside 0.1~au for typical flare stars
(Fig.~\ref{fig:wedge}). If combined with transits and radial velocity
measurements, which provide separate constraints on stellar and
planetary radii and masses, the possibility of using echoes to
constrain the planet's surface composition is greatly enhanced.

The ideal star for planetary echo detection has many strong, short outbursts. 
As a class, M dwarfs seem particularly well suited. 
A subset of these stars (the dMe) have tens of flares per day \citep[e.g.,][]{pettersen1986, hawley2014, davenport2016, yang2017}, and the contrast between the flare emission and the quiescent stellar light can be as high as $10^{4}$--$10^{5}$ \citep[e.g.,][]{schmidt2014, schmidt2016}.  
This combination --- high flare frequency and strong flare contrast --- make these stars, including AD Leonis \citep{bromley1992} and Proxima Cen \citep{sparks2018}, excellent preliminary candidates for light echo detection.
Furthermore, M dwarfs are the most common type of star in the galaxy\citep{ledrew2001}, providing a very large number of potential observation targets.

The detection of faint echoes is an extraordinary challenge, calling for a commitment of both collecting area and observing time.
Observation of echoes from an exoplanet requires (i) analyzing a signal that is typically buried in the measurement noise, (ii) isolating the echo from the possibly complex flare tail and post-flare brightening activity, (iii) accounting for the probability that the exoplanet is in a favorable position relative to both the flare and Earth, and (iv) developing detection significance criteria.  
While the concepts motivating this `interstellar lidar' are straightforward, there is no generalized approach available to enable demonstration.

Here, we propose a framework for exoplanet echo detection by statistical analysis of the autocorrelation signals of multiple high-contrast flare events measured at high cadence (typically 10~s or faster).
Through simulations, we show that we are able to extract echoes and orbital and other parameters by additively building up a signal using data from a sufficiently large number of flares and, in most cases, can place confidence intervals on the data by using statistical analysis and resampling techniques.  
As we show here, even if echoes are not individually resolved in light curves, their signatures can rise above noise when combined together using assumptions about a planet's orbit to account for variations in the echo delay time. 
In practice, confirmation of the presence of faint echoes and estimates of the planet's orbital elements go hand in hand.

We organize this paper as follows.
In \S\ref{sect:method}, we review echo detection and describe the algorithms we use for selecting and processing flare light curves.  
Toward understanding the feasibility of our method, we also provide
examples of successful planet detection and parameter extraction from mock flares. 
Then, in \S\ref{sect:framework}, we provide a framework for using faint
echoes for planet discovery, with details for how to 
explore the inner regions of planetary systems by tracking flare activity.
We consider the prospects of using this method for planet discovery or follow-up observations in \S\ref{sect:prospects} and we conclude in \S\ref{sect:conclude}.

\needspace{6em}
\section{Feasibility of faint echo detection}\label{sect:method}

\begin{figure}
\centerline{\includegraphics[width=6.0in]{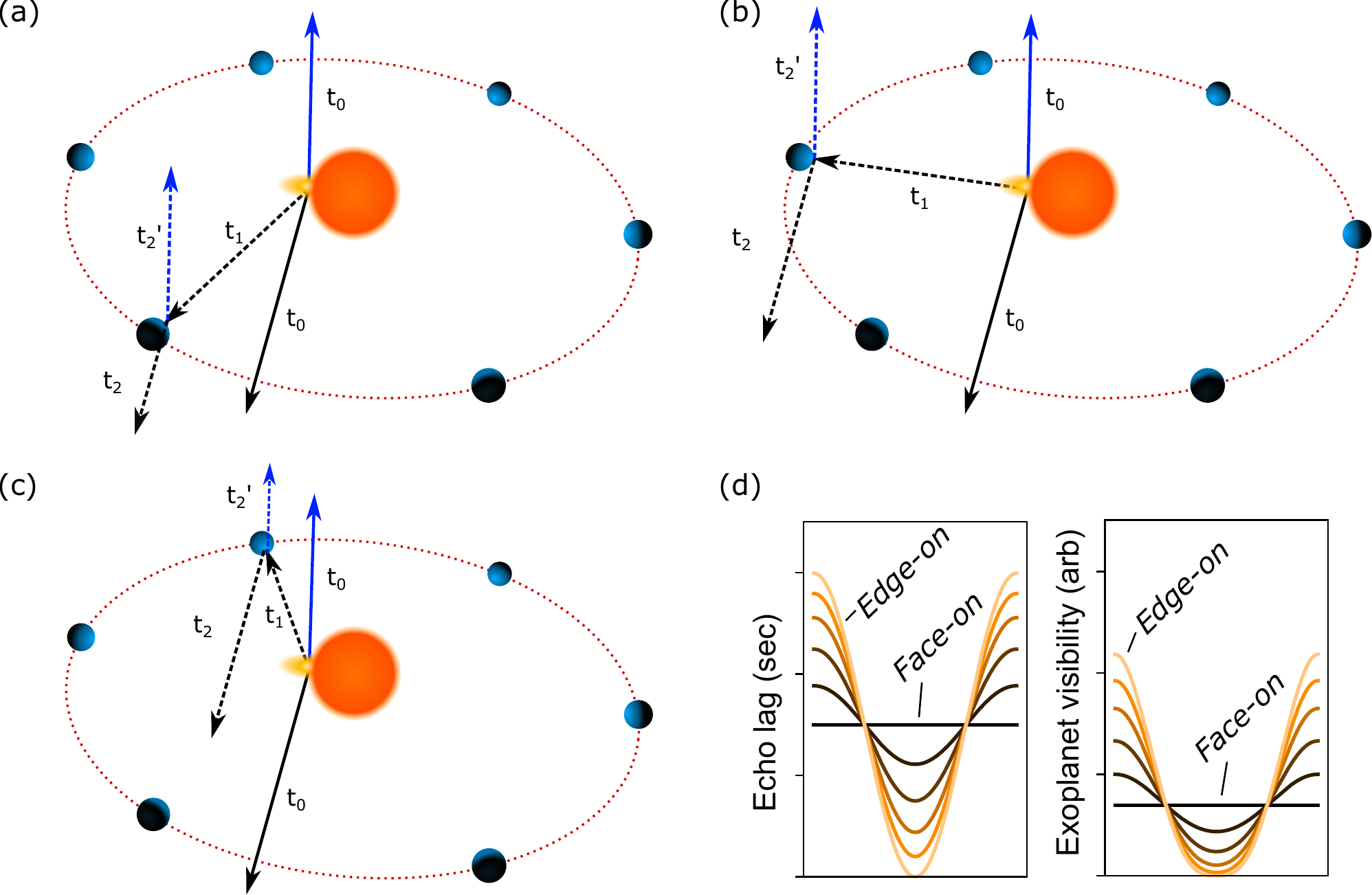}}
\caption{\label{fig:echogeom} Schematic of a circular orbit is depicted at a perspective angle illustrating the orientation-dependent planetary echo from a flare event.
	The solid-line arrows represent the signal from the flare event; the dashed-line arrows represent the signal from the echo.
	The blue-line arrows travel to a face-on observer; the black-line arrows travel to an edge-on observer.
	The light travel time from the star to the exoplanet, $t_1 \sim a/c$, remains constant for all orbital positions.  
	The light travel time from the star to Earth, $t_0$, is assumed the same for all orientations.
	The exoplanet-Earth light travel time, $t_2$, is therefore the only parameter that depends on the orientation.
	The echo lag time, $\tau = t_2 + t_1 - t_0$, depends on the specific orientation.
	(a) When the exoplanet approaches inferior conjunction (the transit configuration), $\tau \rightarrow 0$ as $t_2 \rightarrow t_0-t_1$.  
	(b) When the exoplanet is observed near quadrature, the delay time is the light travel time from the star to the planet, $\tau \sim a/c$ as $t_2 \sim t_0$, which is equivalent for edge-on and face-on orientation.
	(c) When the exoplanet approaches superior conjunction, $\tau \rightarrow 2 a/c$ as $t_2 \rightarrow t_0 + t_1$.
	(d) The relative echo lag and approximate exoplanet visibility as a function of orbital phase and orientation for a circular orbit.
	}
\end{figure}

In a flare event, some photons travel directly to an observer while others
take a different route, scattering off the surface of a planet before heading
to the observer. The difference in the light travel time is the echo delay.
The strength of the echo depends on the relative orientation of the planet to the flare and the observer, as well as the absorption and phase characteristics of the planet.
These two main characteristics of a planetary echo, delay time and strength, are illustrated in Figure~\ref{fig:echogeom}.  
A face-on, circular orbit provides the simplest case with echoes of constant delay time, neglecting the spatial extent of the star and flare, and identical strength, neglecting native variability of the albedo.
An edge-on observation of the same planet has echoes that are strongest when the planet is on the opposite side of the star; the echo delay time is longest there as well, equal to twice the delay time in the face-on case. 
When the planet is in a transiting configuration, there is no echo delay, and the echo strength vanishes. 
Intermediate viewing angles and planets on eccentric orbits give sets of echo time delays and strengths that encode orbital parameters. 
We begin setting up a framework for how to decode this information by examining the nature of individual flares and how to find faint echoes buried in them.

\subsection{Flare Analysis}\label{sect:flares}

Flare events typically exhibit common characteristics \citep{pettersen1986, loyd2014, kowalski2016}: a rapid rise to peak brightness that can be seconds or less, followed by a slower decay characterized by a time $\thalf$, the full-width at half-maximum of the flare, that can range from seconds to minutes. 
Finally, a gradual, more slowly dissipating phase can persist from under a minute to hours. 
For echo detection, it is simplest to work with shorter-lived flares, for which the early, rapidly varying part of the flare light curve vanishes by the time the echo arrives at the detector. 
Echoes from a planet within 0.1~au of the host star have delay times that are typically seconds to minutes, depending on orientation. 
Thus, at least a subset of flares may be well-suited to echo detection.

For echo detection of a planet around a flare star, we first consider an observed light curve for a single flare event. 
To evaluate detection bounds, we assume that the flare is an isotropic point source, impulsive relative to the cadence, appearing in one bin only, producing a signal $\intflarepeak$ (e.g., in units of photon counts) above the constant stellar background rate, $Q$.
The r.m.s. noise level in the continuum is $\rmsnoi$ and $\signoise$ is the minimum signal-to-noise ratio that the astronomer has determined is acceptable for the detection experiment (with the practical constraint $\signoise \geq 1$), giving a detection regime of
\begin{eqnarray}\label{eq:eps}
\epsilon & \gtrsim & \frac{\signoise \rmsnoi}{\intflarepeak} 
\approx
\frac{\signoise \sqrt{Q \deltat}}{\intflarepeak}
\end{eqnarray}
where $\epsilon$ is the exoplanet-star (echo-flare) contrast ratio, the integrated time bin is of duration $\deltat$, and the second line assumes that the noise comes from counting statistics in the continuum integrated in a time bin.  
However, flares typically have a broad tail and cannot be easily summed in a single time bin.

\begin{figure}
\centerline{\includegraphics[width=5.0in]{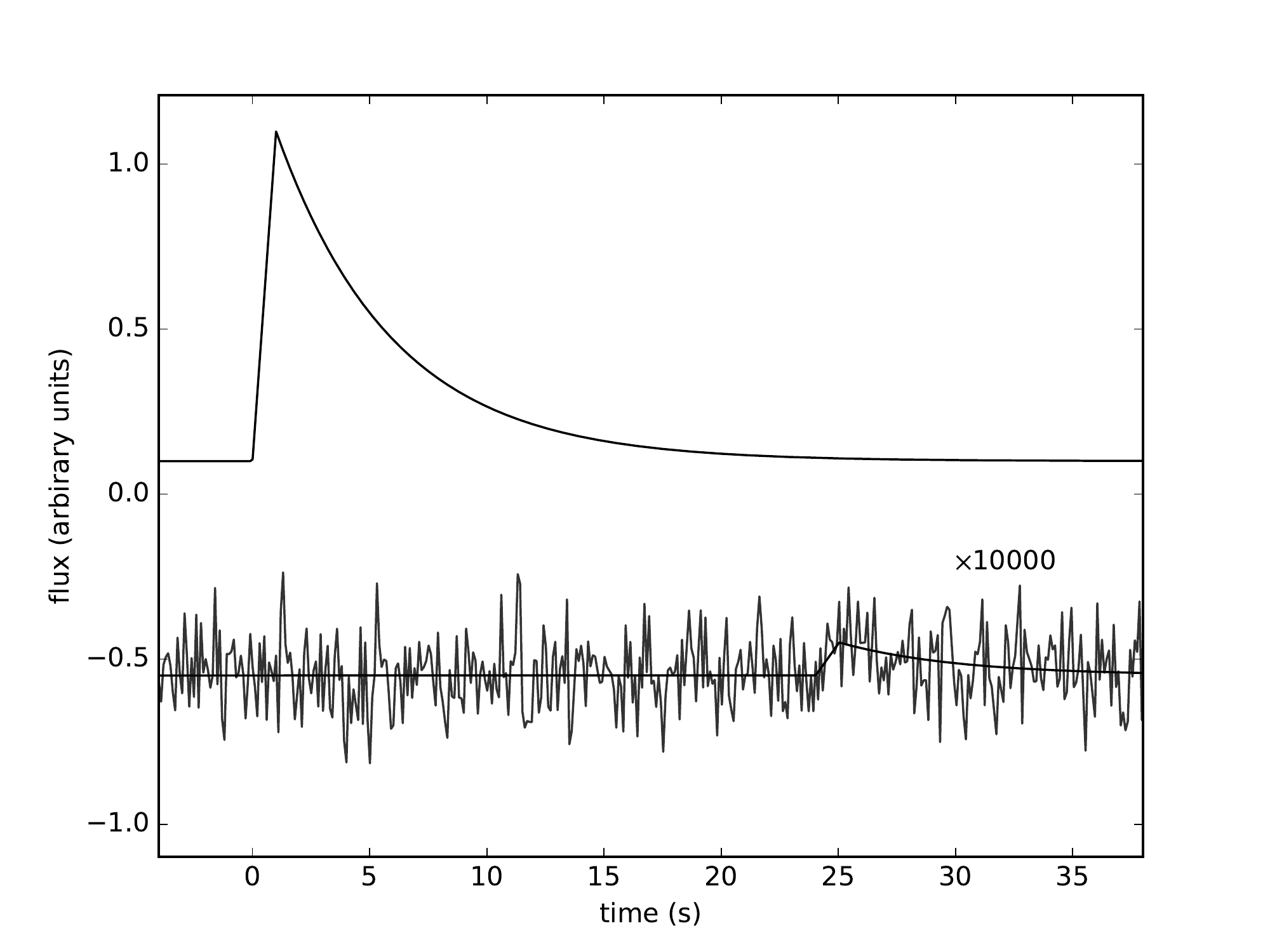}}
\caption{\label{fig:flare} Simulated flare with an echo.  The echo
  amplitude and synthetic Gaussian noise, visible in the lower curve (shifted downward
  and scaled with a multiplicative factor for clarity) are both at the
  level of $10^{-5}$ relative to the flare amplitude.}
\end{figure}

To provide a method for more realistic flares, we consider the typical structure of a flare measurement.
The light curve, $L(t)$ (in units of photon counts per second, typically), has contributions from a single flare with flux $F(t)$, its echo, and measurement noise $N(t)$:
\begin{equation}\label{eq:lightcurve}
L(t) = Q + F(t) + \epsilon F(t-\techo) + N(t),
\end{equation}
where $\epsilon=p \phi(\alpha)(R^2/a^2)$ is the approximate echo contrast of an exoplanet with semi-major axis $a$, geometric albedo $p$, radius $R$, and phase function $\phi(\alpha)$, and $\techo$ is the echo delay, both of which depend on the geometry of the star (where the flare presumably originates), the planet (responsible for the echo), and the position of the observer.
Figure~\ref{fig:flare} illustrates an idealized flare light curve with an echo whose strength ($\epsilon = 10^{-5}$) is comparable to the noise level.

The extraction of the faint echo in the presence of noise is a problem in signal processing.  
Since a flare echo is just the time-delayed repetition of a flare, a good tool for extracting the echo is the autocorrelation function \citep{argyle1974, bromley1992}. 
This statistical tool provides a sensitive measure of the similarity of a given signal to itself as a function of time delay. 
A detailed description of the autocorrelation analysis is provided in Appendix~\ref{appx:autocorr}.  

We caution that, even in ideal scenarios with large echo contrast, it may be difficult to discern a flare with a planetary echo from a multi-peaked flare. 
The autocorrelation function might not yield a distinction between the two cases. 
Furthermore, a flare may not produce an echo at all, as when a flare is localized on the stellar surface, but on the opposite side relative to the planet.
These challenges associated with realistic observing conditions are considered in the following sections.

\subsection{Observational requirements for echo detection}\label{sect:feasible}

To provide meaningful detection criteria, even in scenarios with large signal-to-noise ratios, multiple flares are needed to rule out false alarms like multiple peaks in the same flare event.  
When a planet is on a circular orbit and observed face-on, all echoes occur with the same delay time, and so add together to strengthen the signal. 
In general, the planet's orbital motion causes echoes in a set of flares to occur at different times, but we can compensate for this effect so that echo signals in the correlation function are aligned, adding together at some common lag (see \S\ref{sect:lags} below). 
Then, the total echo signal grows linearly with the number of flares, $\nflares$, while the noise increases as the square root of this number.
This linear dependence may change depending on the flare brightness distribution; see for example \citep{sparks2018}.
Thus, the typical echo contrast required for detection falls off with $\nflares$ as
\begin{eqnarray}\label{eq:epsnflares}
\epsilon 
& \gtrsim & \frac{\signoise}{\langle \intflarepeak \rangle} \sqrt{\frac{Q \deltat}{\nflares}},
\end{eqnarray}
which follows from Eq.~\ref{eq:eps}, with $\langle \intflarepeak \rangle$ as the average flare strength. 
To obtain approximate observation times required to detect an exoplanet, we define $N_{f}=\nu_{f} t_{obs}$, where $\nu_{f}$ is the flare frequency and $t_{obs}$ the total observation time and consider a Lambertian planet phase function\citep{traub2010} to obtain:
\begin{equation}\label{eq:detectiontime}
t_{obs} \gtrsim \frac{\signoise^2 Q \deltat}{\nu_{f} \langle \intflarepeak \rangle^2 \epsilon^2} = \frac{\signoise^2 Q \deltat}{\nu_{f} \langle \intflarepeak \rangle^2} \frac{\pi^2 a^4}{p^2 R^4}
\end{equation}
where the $\pi^2$ term arises from viewing the planet from a face-on configuration, see Eq.~\ref{eq:lambertian}.
In reality, this factor may need to be multiplied by 2 to account for half the flares missing the exoplanet, presumably occurring on the opposite side of the star from the planet, but still being observed by our telescope.
From Equation~(\ref{eq:detectiontime}), we see that the time required to monitor the star goes as the inverse square of the exoplanet-star contrast, $\epsilon$, which goes as the inverse square of the distance from the star.
Hence, the monitoring time depends on the star-planet separation to the fourth power. 
Therefore, stellar flare echoes are best detected from exoplanets that are near to their host star.

\begin{figure}
	\centering
	\subfloat[Simulated light curve with and without noise.]
	{\includegraphics[width=.55\textwidth]{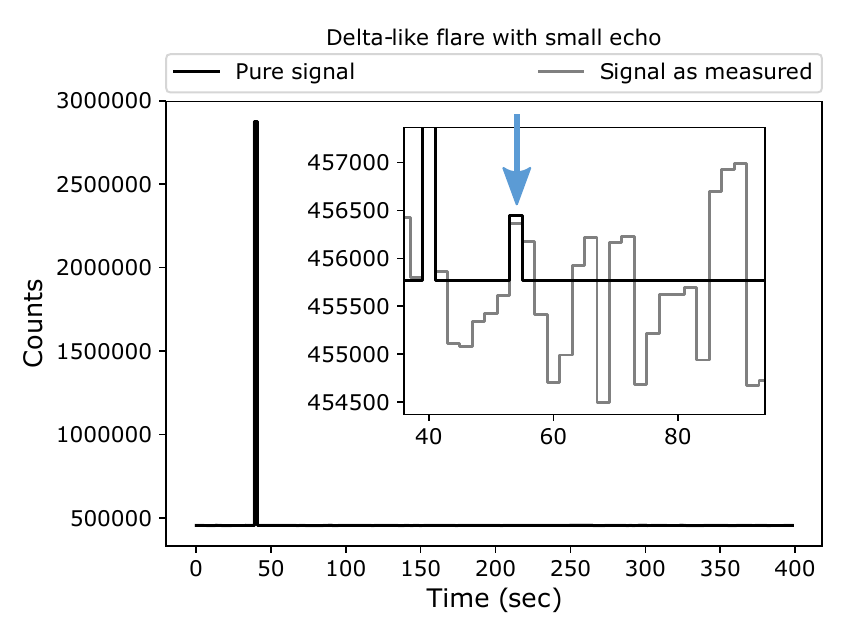}\label{fig:light_curve1}}
\hfill
	\centering
	\subfloat[Detrended correlation function for increasing numbers of flares.]
	{\includegraphics[width=.425\textwidth]{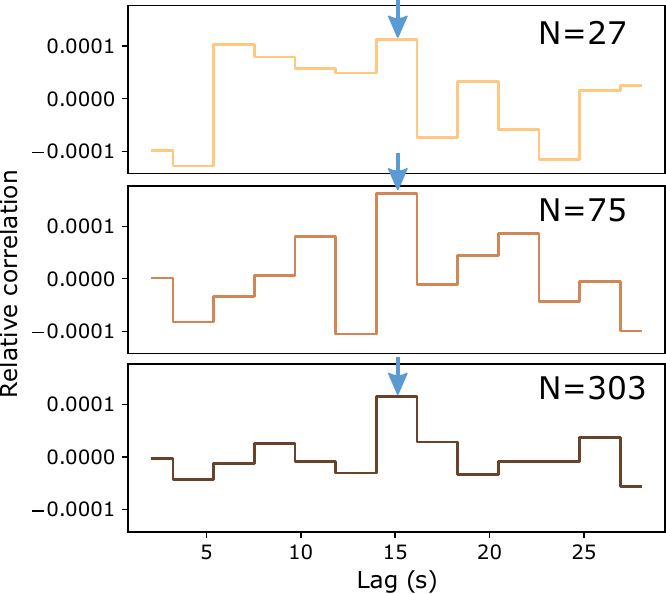}
	\label{fig:convergence}}
\caption{The light curve and correlation function for ideal 
impulsive flares from a star with $m_{U}=12$, with flare magnitude $\Delta m_{U}=-2$, observed by a 2-m diameter telescope with a 2~s integration time.  
An echo comes from a planet at $a = 0.03$~au, on face-on circular orbit, with $R=2R_{J}$ and  $p=0.9$. 
The resulting echo delay time is 15 seconds.
The flares are identical, but the noise profiles are generated independently for each sample.  
The plots illustrate how multiple flare events are essential for faint echoes to produce a significant imprint in the correlation function. With 27 flares, the echo corresponds to an insignificant peak in the correlation function. When 75 flares are included, the echo is the largest peak in the correlation function, but it is of similar magnitude to some other false peaks.}
\label{fig:delta_light_curve}
\end{figure}

To demonstrate the feasibility of using this technique under ideal conditions, we produce a set of proof-of-concept simulations. 
First, we consider impulse flares that occur within a single time bin, and set 50\% of the flares to miss the exoplanet, which systematically de-correlates the signal.  
We select U-band magnitude $m_{U}=12$ with flares that result in $\Delta m_{U}=-2$, which are parameters relevant to many M dwarfs.
We then generate an echo consistent with a hot Jupiter at 0.03~au, $R=2R_{J}$, albedo of 0.9, viewed in a face-on circular orbital configuration. 
To quantify the results, we calculate the signal that would be observed with a 2~m diameter telescope with a 2~s integration time and then add counting noise with the numpy random.poisson function\citep{numpy2006}.
The flares are assumed to be non-overlapping in time.
The resulting data have an echo-magnitude/mean counting-noise (signal-to-noise) ratio of $s=1.0$ for each flare.
Under these conditions, using Eq.~\ref{eq:detectiontime}, the predicted number of flares required to achieve $s=\{3, 5, 10\}$ is $\{27, 75, 303\}$.
Figure~\ref{fig:delta_light_curve} provides an example of a simulated light curve from this experiment with and without noise and its detrended correlation function for increasing numbers of flares.

Under real observing conditions, the nearly face-on circular orbit case appears to be the most promising configuration because all echoes will accumulate within one or two time bins, avoiding the need to align echo lags.
Unfortunately, a signal accumulating within a common lag bin is also a noise condition that can be expected from real instrumentation\citep{mann2017}.
Therefore, using these data sets, we (i) analyze the direct autocorrelation signal, and (ii) perform a statistical hypothesis testing experiment that is motivated by the system phenomenology.
Because half of the flares are hidden from the hypothetical planet and will not produce echoes, we would expect a bimodal distribution of the correlation values at the time of the lag, and a unimodal distribution elsewhere when there is no echo.  
These two cases, a null hypothesis of no echo with a unimodal distribution and a positive detection hypothesis of a bimodal distribution, are now open to the vast toolkits available from statistical analysis.
For instance, with a sufficiently large dataset, the detection case will typically manifest itself as having a larger variance and smaller kurtosis and discernible differences in the kernel density estimation.  

\begin{figure}
\centerline{\includegraphics[width=.85\textwidth]{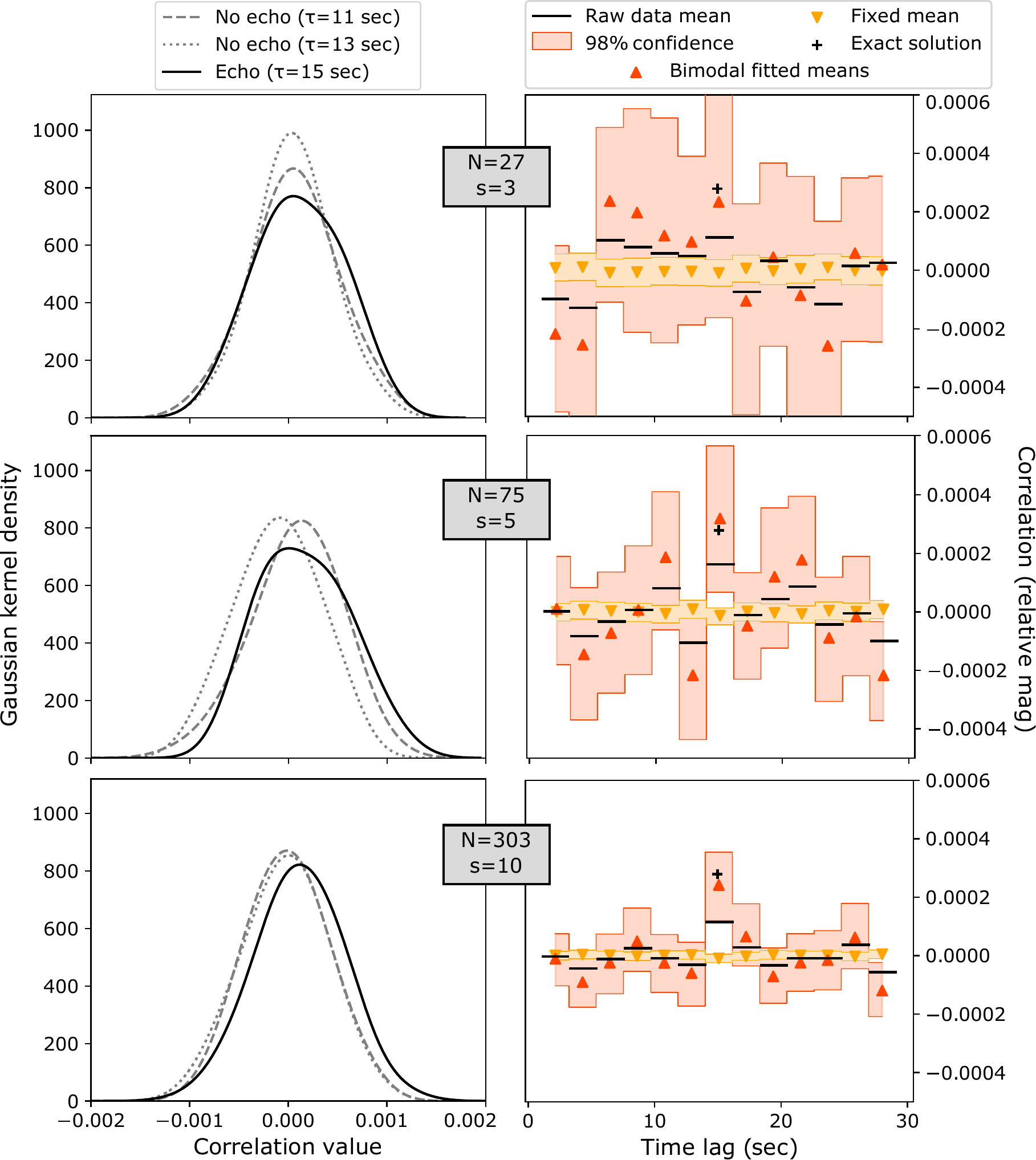}}
\caption{(Left) Gaussian kernel density estimates at the time of the echo and at two times without echoes. (Right) Bi-Gaussian fits of the correlation data for $N=\{27, 75, 303\}$ flares with $R=2R_{J}$ (same dataset shown in Fig.~\ref{fig:delta_light_curve}).  
The bimodal model consists of two identical Gaussian profiles, where one is fixed at a location consistent with no echo signal, and the other is allowed to float. 
While visibly subtle, the bimodality of the kernel density estimate proved to be quantitatively robust under bootstrap analysis.
The location of the floating Gaussian mean becomes distinct from the fixed one at better than 98\% confidence level at the lag equal to the echo delay time in the cases with $N=75$ and $303$.
The parameter $s \sim \sqrt{N}$ indicates the strength of the echo contributions in each data set.
}
\label{fig:confidence}
\end{figure}

In Figure~\ref{fig:delta_light_curve}, we examine direct autocorrelation echo detection in the model where a planet is observed face-on, using data sets with 27, 75, and 303 mock flares. 
We find that the autocorrelation time lag corresponding to the echo, indicated with an arrow, produces a strong signal, but is not convincingly distinct from the background until the 303 flare case.
Then, we explore a hypothesis testing analysis: instead of summing the autocorrelation values at each time lag, we compute kernel density estimates of the autocorrelation functions at each time lag, using a normal kernel with a bandwidth determined by Silverman's (\citeyear{silverman1986}) rule of thumb.\footnote{This is a simple estimate; a more accurate bandwidth can be produced by the evaluating the sensitivity index, per Eq.~\ref{eq:detectionsensitivity}.}
The left side of Figure~\ref{fig:confidence} shows these estimates at the time lag equal to the echo delay time (15~s), and at two different lags for which the echo signal would not contribute to the autocorrelation function.
With $N=27$ mock flares contributing to the correlation function, the echo case is not meaningfully discernible from the no-echo case.
For the $N=75$ and $N=303$ cases, there is a difference in the mean of the distributions and the shape of the tail of the kernel density estimate.

From the kernel density estimate shown in Fig.~\ref{fig:confidence}, we calculate the cumulative distribution function and fit it with the cumulative distribution function of a bi-Gaussian model where both Gaussians are assigned the same amplitude and width, consistent with the model's anticipated phenomenology.
In each case, one Gaussian's mean is allowed to float, while the other is 'fixed' at the mean of the correlation values in adjacent time-lag bins.
In more general models, the fixed position can be selected from a detrended background curve, for example.
The model fit thus has two free parameters: a Gaussian mean and the width of both Gaussians.
We perform the fitting with the SciPy optimize.curve\_fit routine\citep{scipy2001}.
The fitted mean position at each time lag is shown in the plots on the right side of Fig.~\ref{fig:confidence} as orange triangles and the positions of the fixed Gaussian is shown as an upside down triangle.
These values are then compared to the raw autocorrelation values, shown as a horizontal black bar and taken from the data in Fig.~\ref{fig:delta_light_curve}.

Toward producing meaningful confidence limits, we resample the correlation values with replacement (a statistical method known as bootstrapping) and re-calculate the two-Gaussian model fit for each resampled distribution.  
We use 1000 resampled sets and evaluate the upper and lower confidence intervals with Python's numpy percentile function.\citep{numpy2006}\footnote{Note: we do not anticipate this data to have a Gaussian distribution, so we explicitly sampled the upper and lower values from the bootstrap-resampled values.}
The light orange bars in Figure~\ref{fig:confidence} show the position of the 98\% confidence values.
When the bars from the floating Gaussian mean overlaps the bars from the fixed Gaussian mean, we say that there in insufficient information to claim a detection event.
This is the case for all time lags in the N=27 case.
When a larger number of flares are observed, the 98\% confidence interval of the fitted Gaussian is far enough away from those of the fixed Gaussian such that there is no overlap between the two at only one time, 15~s.
Here, we can say that this model indicates an echo detection event with 98\% confidence.
This occurs in the N=75 case and is more significant in the N=303 case.
These detection events correctly correspond to the time and magnitude of the synthetic echo; the fitted Gaussian mean is close to the exact solution (plus sign in Figure~\ref{fig:confidence}).

Given the high flare rates for M dwarfs, it should be plausible to select only the most impulsive flares and use them for such an analysis.  
For many stars, this will result in more than one useful datum per day, though additional detailed high-cadence measurements are necessary to say quantitatively how many useful flares can be expected for a given star class.

\subsection{Echo detection with planetary motion}\label{sect:lags}

In the more general configurations when a planet's orbit is not viewed face-on, the echo delay times may be different for each flare. 
If the echoes are buried in noise, then their signals cannot be fruitfully combined unless the true echo delay times are known. 
We can make progress by modeling the planet's orbit and calculating echo delays accordingly.
The relevant parameters are the host star mass, $M$, and the planet's Keplerian orbital elements $a$, $e$, and $M_0$, where $a$ is the semimajor axis, $e$ is the orbital eccentricity, and $M_0$ is the mean anomaly at some reference time $T_0$.  
The echo delay will also depend on spherical polar viewing angles $\theta_v$ and $\phi_v$, such that $\theta_v = 0$ mean a face-on view, while $\theta_v = 90^\circ$ and $\phi_v = 0$ means the planet is transiting the star at periastron.
With these parameters (six in total, counting the stellar mass), and assuming a point-like star, we can calculate an echo delay time for any orientation and orbital phase using the equation for a constant-delay ellipsoid, borrowed from radar theory:
\begin{equation}\label{eq:echoorbit}
    \techo = \left(|\vec{r}| -
    \vec{r}\cdot\hat{e}_\text{obs}\right)/c,
\end{equation}
where $\vec{r}$ is the position of the planet at the time of a flare, $\hat{e}_\text{obs}(\theta_v,\phi_v)$ is a unit vector pointing at the observer from the host star, and $c$ is the speed of light. 

\begin{figure}
\centerline{\includegraphics[width=6in]{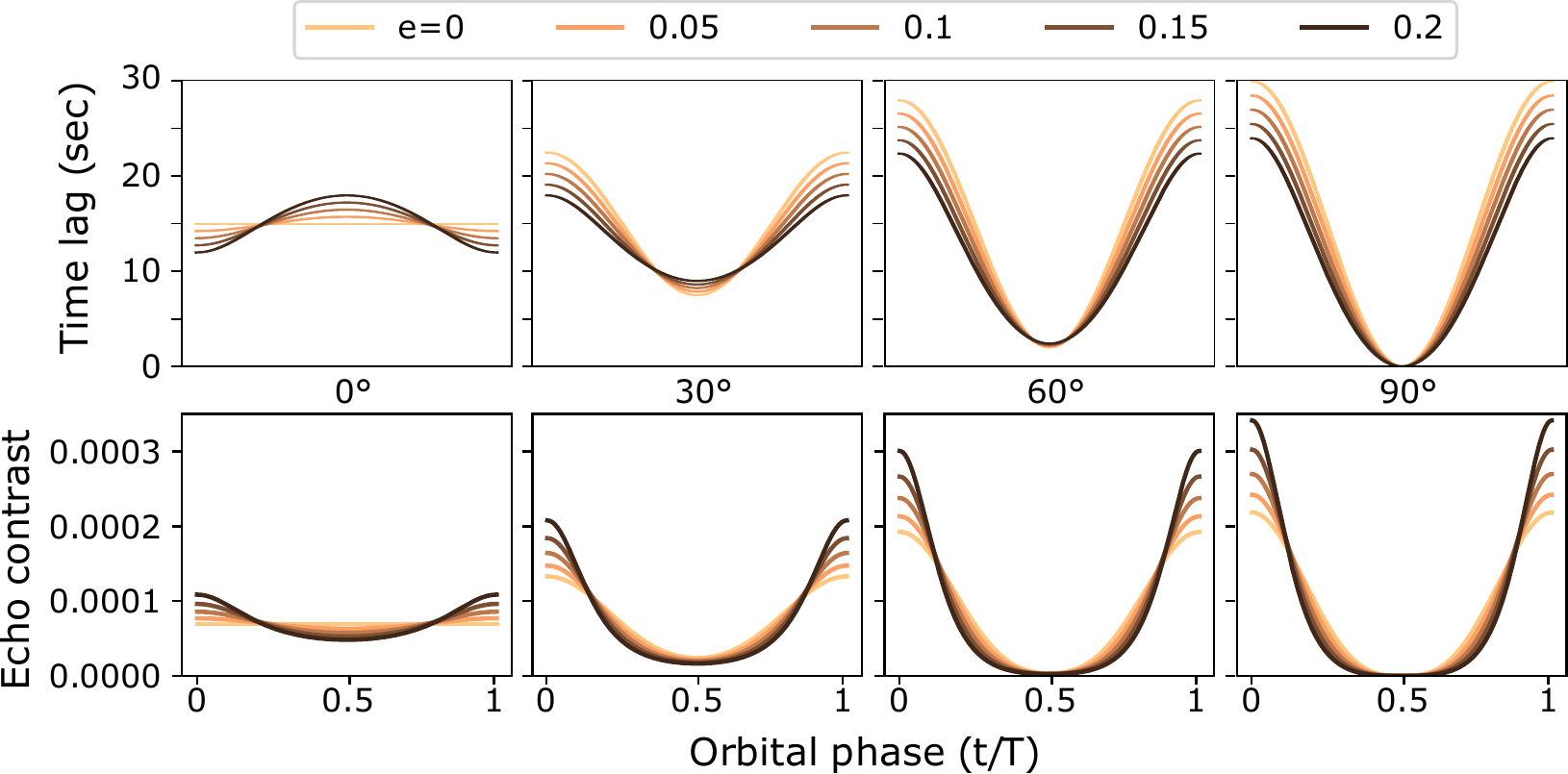}}
\caption{\label{fig:orbitlags}
Echo delay time lag\citep{mann2017, sparks2018} (top row) and star-exoplanet contrast ratio observed from Earth (bottom row) from a planet with $R=2R_J$, albedo of 0.9 at 0.03~au for eccentricities $\{0, 0.05, 0.1, 0.15, 0.2\}$ (darker lines correspond to higher eccentricities) for different inclination angles relative to Earth (face-on view is on the left, transit view is on the right). 
In these scenarios the viewer is nearest to the planet's apoastron location.
Eccentricity plays a more significant role for the nearly face-on configurations ($0-30^\circ$) where the differences in the exoplanet-star separation drives differences in the echo lag.
For more extreme inclination angles, ($30-90^\circ$) the viewing angle drives most of the variation in both the time lag of an echo following a flare, and the star-exoplanet contrast.
For example, in the face-on case, the the echo time lag is maximized when the exoplanet is furthest from the star ($t/T=0.5$ in this figure), but by a $30^\circ$ inclination angle, the time lag is actually minimized when the exoplanet is furthest from the star.  
This peculiarity arises because of the specific viewing angle relative to the planet's apoastron location: despite being at its furthest distance from the star, it aligns closer and closer to the view of Earth with increasing viewing angle, reducing the echo lag to zero.
Note that, using a combination of both lag time and echo contrast, we can distinguish between each case, despite the similarities of individual curves as a function of orbital phase.
}
\end{figure}

Figure~\ref{fig:orbitlags} shows examples of how echo delay times and echo strengths depend on the orbital phase for various eccentricities and viewing angles. 
For the modest eccentricities considered here ($e\leq 0.2$), changes in viewing angle produce the most significant variations in delay times and echo strength. 
For example, a planet viewed nearly edge-on and from the opposite side of the star relative to the observer produces a strong echo with a large delay (twice that of a face-on view). 
As the planet moves between the observer and the star, the delay time decreases, and the echo strength drops, like the brightness of a waning moon.
We show several more examples in Appendix~\ref{appx:orbitalsignals}.

\begin{figure}
\centerline{\includegraphics[width=4.5in]{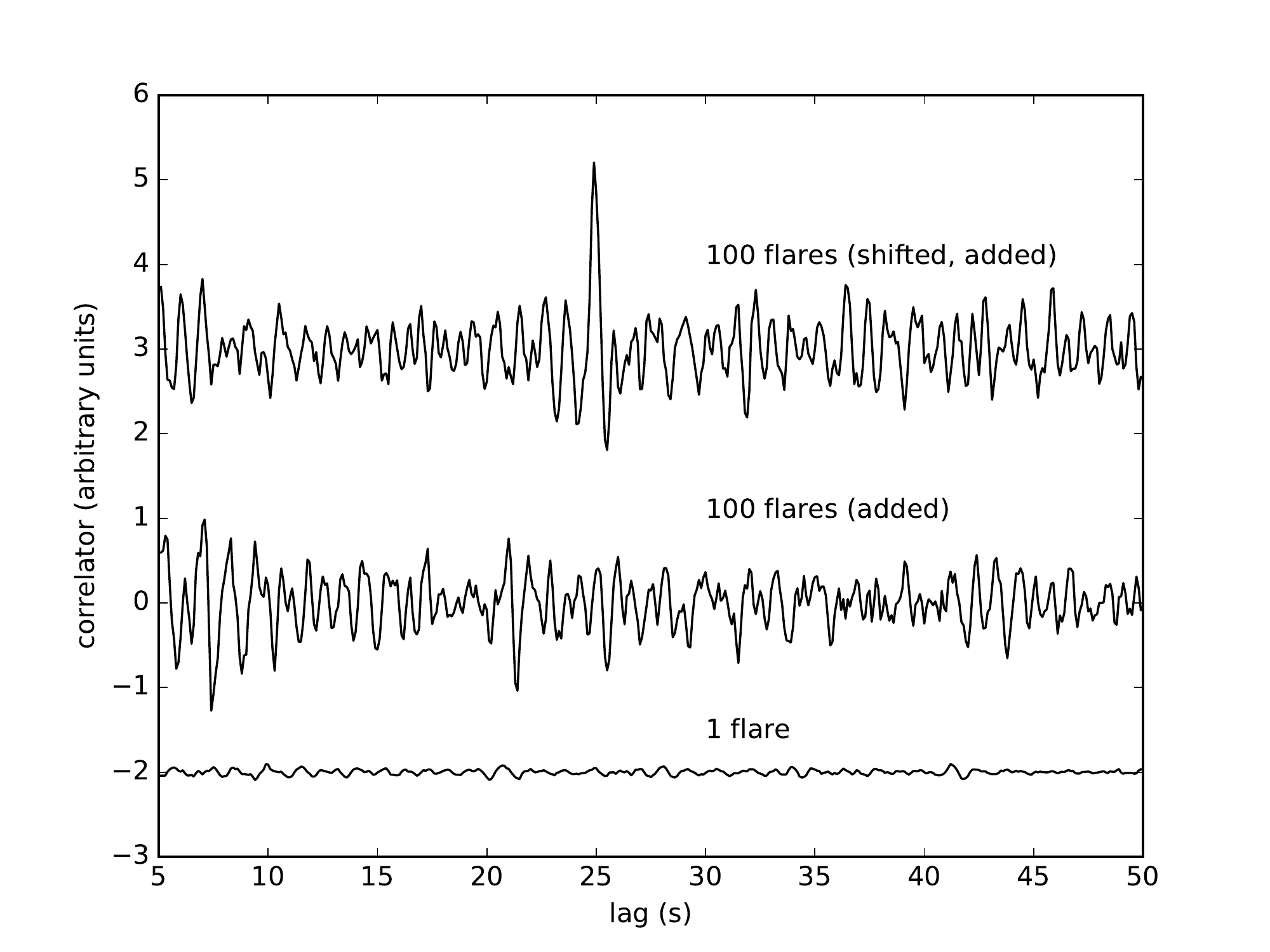}}
\caption{\label{fig:corr} The correlation signal of a set of mock flare echoes, similar to the one shown in Fig.~\ref{fig:flare}, from a planet on a close, mildly eccentric orbit ($a = 0.05$~au and $e = 0.1$).
The mock dataset contains 100 flares, all with echoes. 
The lower curve (offset vertically for clarity) is the correlator of a single flare; no trace of an echo is visible. 
The middle curve is the sum of all 100 correlators, which also shows little sign of any echoes. 
The strong peak in the upper correlator arises, despite that the echoes are buried in the individual light curves, because the lags of the individual correlators are adjusted to align the echo peaks at a lag equal to the average echo delay time ($a/c$). 
Thus, it is essential to compensate for the variation in delay time that arises from planetary motion.}
\end{figure}

Figure~\ref{fig:corr} provides an example of the autocorrelation function of a mock impulsive flare with an echo (see Fig.~\ref{fig:flare}), along with sums of such functions from multiple flares with echoes from a planet on a Keplerian orbit. 
We pick the stellar mass, orbital elements, and viewing angles to calculate echo delays ($M = 0.3\,\Msolar$, $a = 0.05$~au, $e = 0.1$ and $M_0 = 70^\circ$ (relative to a fiducial starting time), $\theta_v = 20^\circ$, with $\phi_v = 60^\circ$). 
In our mock data, there are 100 flares, with onset times drawn at random over an interval that spans 10 planetary orbits, about 75 days.  
For each flare, we assume a linear rise in luminosity, and an exponential tail with a $\thalf$ lognormally distributed about 3~s (e.g., Fig.~\ref{fig:flare}); the rise times are 20\% of the decay time up to a maximum of 1~s. 
In the mock data used to make Figure~\ref{fig:corr}, the planet is moving, hence the echo lags are all different.
Adding the raw autocorrelation functions together fails to reveal a compelling echo signature.  
When the autocorrelation functions are shifted so that the echo signals are aligned at a common lag, a strong peak emerges.
In this flare catalog, the intrinsic flare brightness and the echo strength are held fixed from flare to flare, every flare lights up the exoplanet, and each flare is weighted equally.
We then add noise in the form of independent Gaussian random variables, with the goal of mimicking counting noise. 
Both the echo strength and the r.m.s. noise are at a level of $10^{-5}$ relative to the flare amplitude.
The purpose of these simulations is to test the feasibility of detection algorithms.  
To produce a more realistic catalog of flares, it would be required to include variability of flare amplitudes and phase-dependent planet albedo.\citep{russell1916}

With sufficiently long observations with adequately high
signal-to-noise, we have found that periodogram analysis, such as
period folding and Lomb-Scargle plots, can isolate the echo signal
(see Appendix~\ref{appx:orbitalsignals}).  To be effective, however, 
these approaches require sufficient signal within a single lag bin 
to rise above the noise.  For weaker signals, it is necessary
to align the echo correlator lag to build up an echo signature.  So we
must systematically search through echo delay parameters to compensate
for a planet's motion.  We return to this problem below.  Meanwhile,
we consider how the planet's orbit affects the nature of echoes that
may be observed.

To proceed beyond the circular orbit problem, we require the six parameters listed above to determine echo delays from a planet in Keplerian motion.  
Since we do not, in general, have \textit{a priori} knowledge of these parameters --- even the stellar mass may not be well constrained --- we must search for them somehow, a task that is computationally daunting. 
Thus, we briefly explore a reduced parameterization of an echo delay by assuming that it varies sinusoidally in time (cf.\ Eq.~\ref{eq:lagt}):
\begin{equation}\label{eq:sin}
\techo = A \sin\big(2\pi \frac{t}{T} - P\big) + D,
\end{equation}
where amplitude $A$, period $T$, phase $P$, and mean delay time $D$ are constants. 
This assumption is motivated by the sine-like structure of the lag-orbit plot for many orbital configurations (see Fig.~\ref{fig:orbitlags} and Appendix~\ref{appx:orbitalsignals} for several examples).
The advantage of this parameterization is that the search space is 4D. 
Furthermore, we can substantially limit the search space since some of these parameters are related through physical conditions: 
For example, $A\leq D$, which intuitively means that the echo must always occur after the flare.
Also $D$ (related to the mean distance to the planet) and $T$ (the orbital period) are related, so that one may be constrained by the other using Kepler's third law, along with the uncertainty of the stellar mass.

\begin{figure}[ht]
\centerline{\includegraphics[width=3.75in]{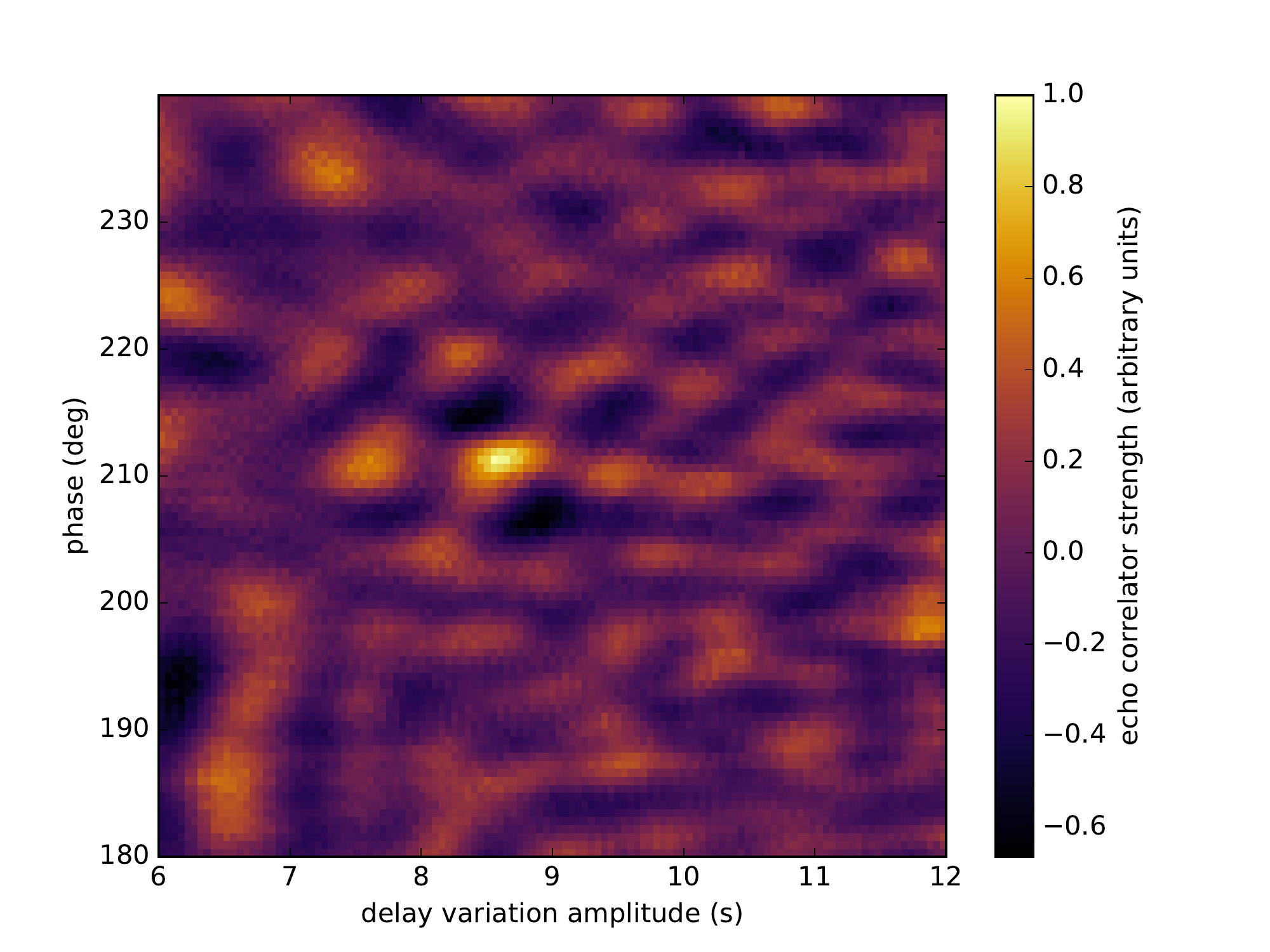}
\hspace{-0.25in}
\includegraphics[width=3.75in]{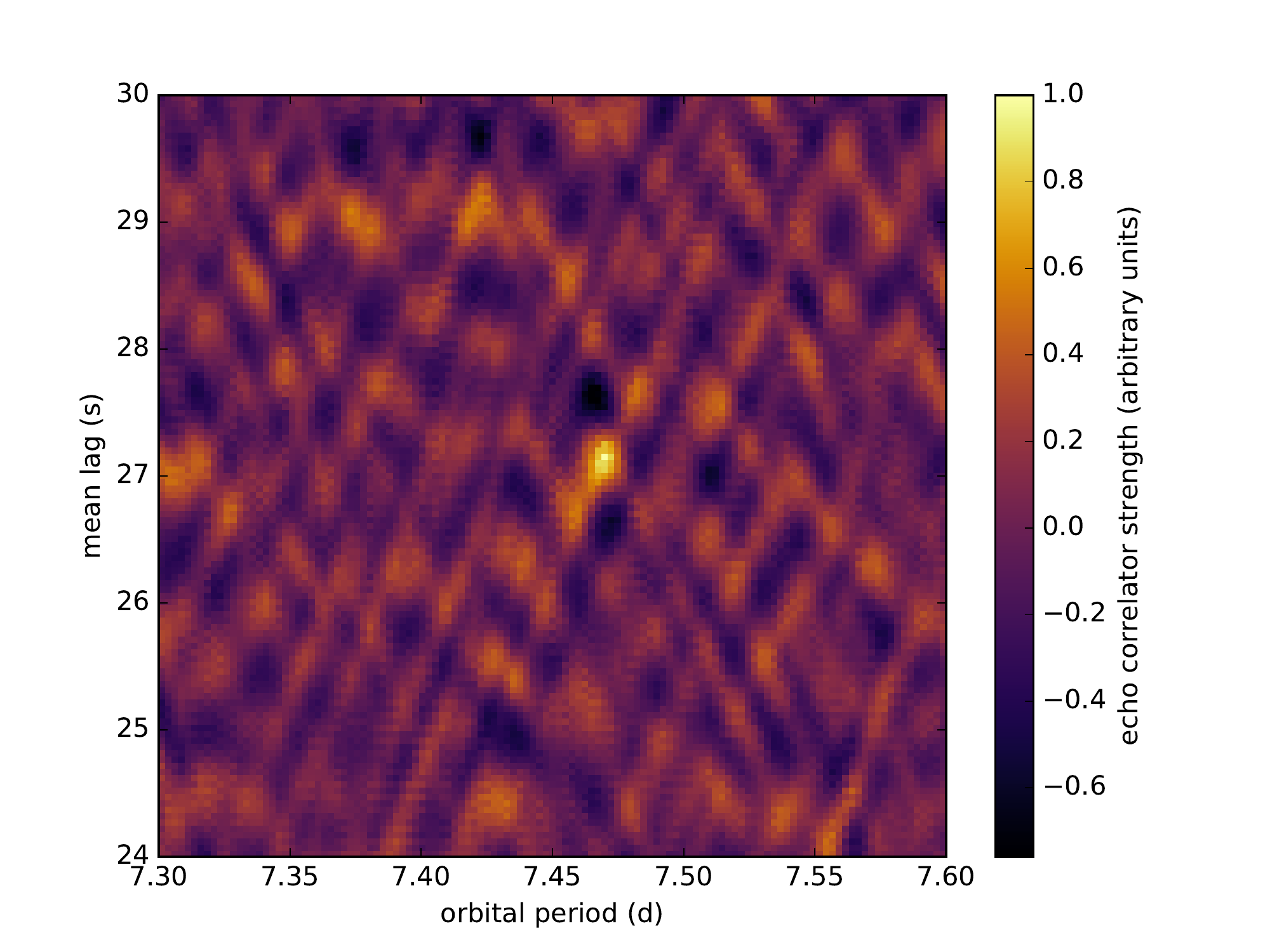}}
\caption{\label{fig:fourdim}Maps of the echo correlator strength ($\mathcal{S}$) from a set of simulated flares (see text and Fig.~\ref{fig:corr}). 
Each panel is a slice of a 4D parameter space that contains the peak value of $\mathcal{S}$.
The left plot is the echo correlator strength as a function of delay variation amplitude ($A$) and phase ($P$), while the right plot shows variations of $\mathcal{S}$ with orbital period ($T$) and the mean lag/echo delay time ($D$). 
The peak of $\mathcal{S}$ lands very close to the expected values $(A,T,P,D)\approx (8.5\ \text{s}, 7.5\ \text{s},
  210^\circ, 25\ \text{s})$.}
\end{figure}

Figure~\ref{fig:fourdim} illustrates the 4D search where we align the 100 individual correlation functions so that the peaks of the correlators would match if the echo delay were given by equation~(\ref{eq:sin}) for a given choice of $(A,T,P,D)$. 
Because a goal will be to work with noisy data such that the echo signal-to-noise is unity or less, we do not just add the aligned correlation functions. 
Instead, we apply a measure --- the echo correlator strength $\mathcal{S}$ --- that responds to a peak at the right lag, \textit{and} the expected shape of the correlation function for each flare echo in the vicinity of the peak (which we know from the shape of the flare's correlation function near zero lag). 
We define one prescription for $\mathcal{S}$ in Appendix~\ref{appx:alg}; We discuss several other possibilities below.
Figure~\ref{fig:fourdim} maps out $\mathcal{S}$ in slices in the parameter space that contains the peak value. 
The strongest peak is close to the expected values, validating the approach.
Once a strong value of the echo correlator strength, $\mathcal{S}$, has been determined, it is then possible to apply a statistical analysis, such as that from \S\ref{sect:feasible}, to produce confidence intervals.

The next step is to generalize our search to any orbital configuration. 
We begin with the stellar mass ($M$) as a parameter. This is needed to relate the orbital period and the orbital distance, which are otherwise formally distinct in the calculation of the echo delay. 
We determine the range of mass values to search from the star's spectral type by assuming an uncertainty of 10\% in $M$. 
If the star was observed to have a bound stellar or planetary partner, we would adopt the dynamical mass estimate instead. 
For each mass choice, we allow semimajor axis, eccentricity, and viewing angles to vary. We focus on inner solar system planets, so we take $a < 0.1$~au, $e < 0.2$ as suggested by \cite{exoplanets}.  
We search the full range of possible orbital phase and viewing angles. 
Our implementation of the method uses a mixture of Python scripts and parallel C++/MPI code. 
The former does the data processing, while the latter performs the grid-based search of billions of points.

\begin{figure}[ht]
\centerline{\includegraphics[width=3.5in]{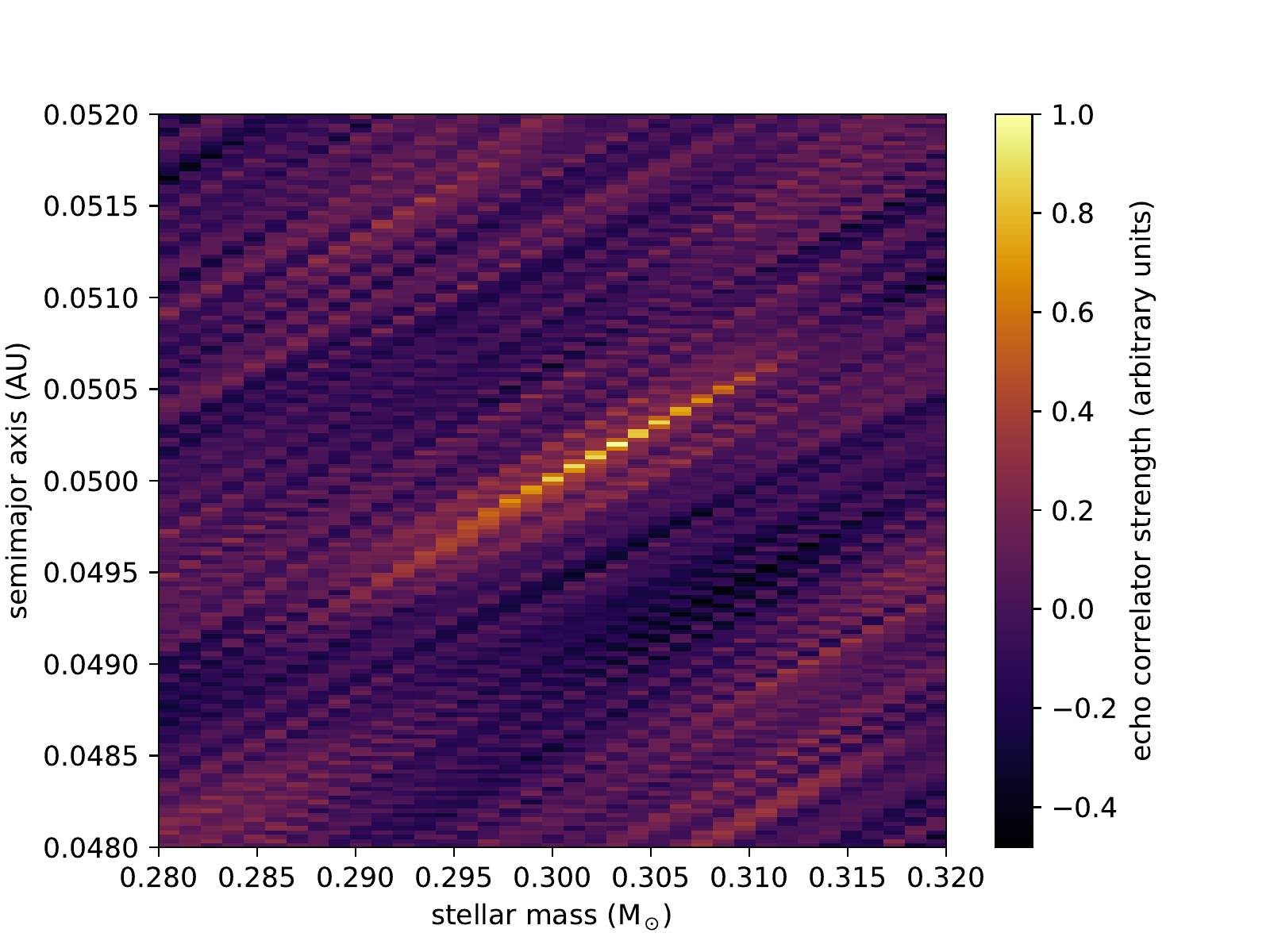}
\hspace{-0.25in}
\includegraphics[width=3.5in]{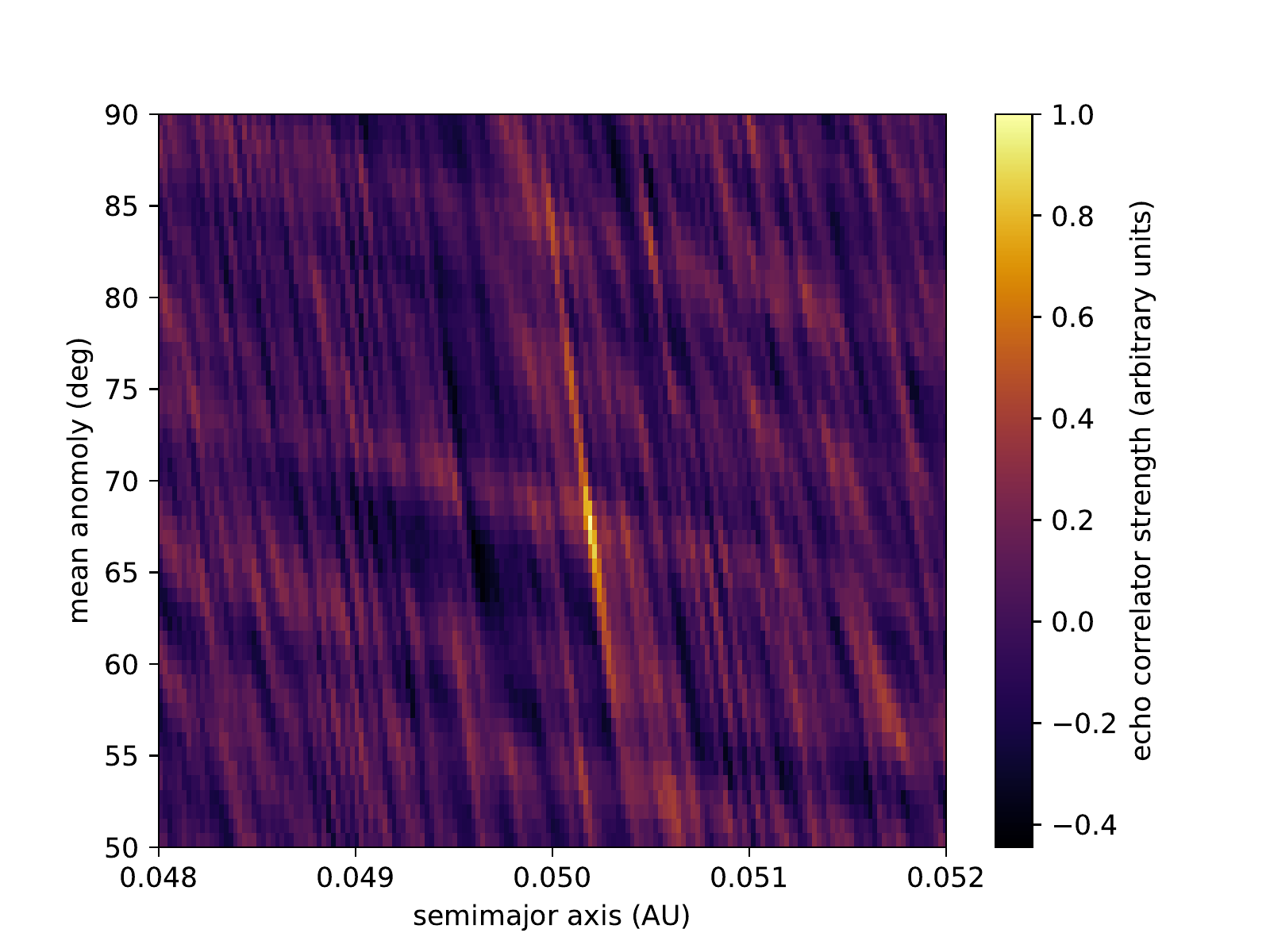}}
\centerline{\includegraphics[width=3.5in]{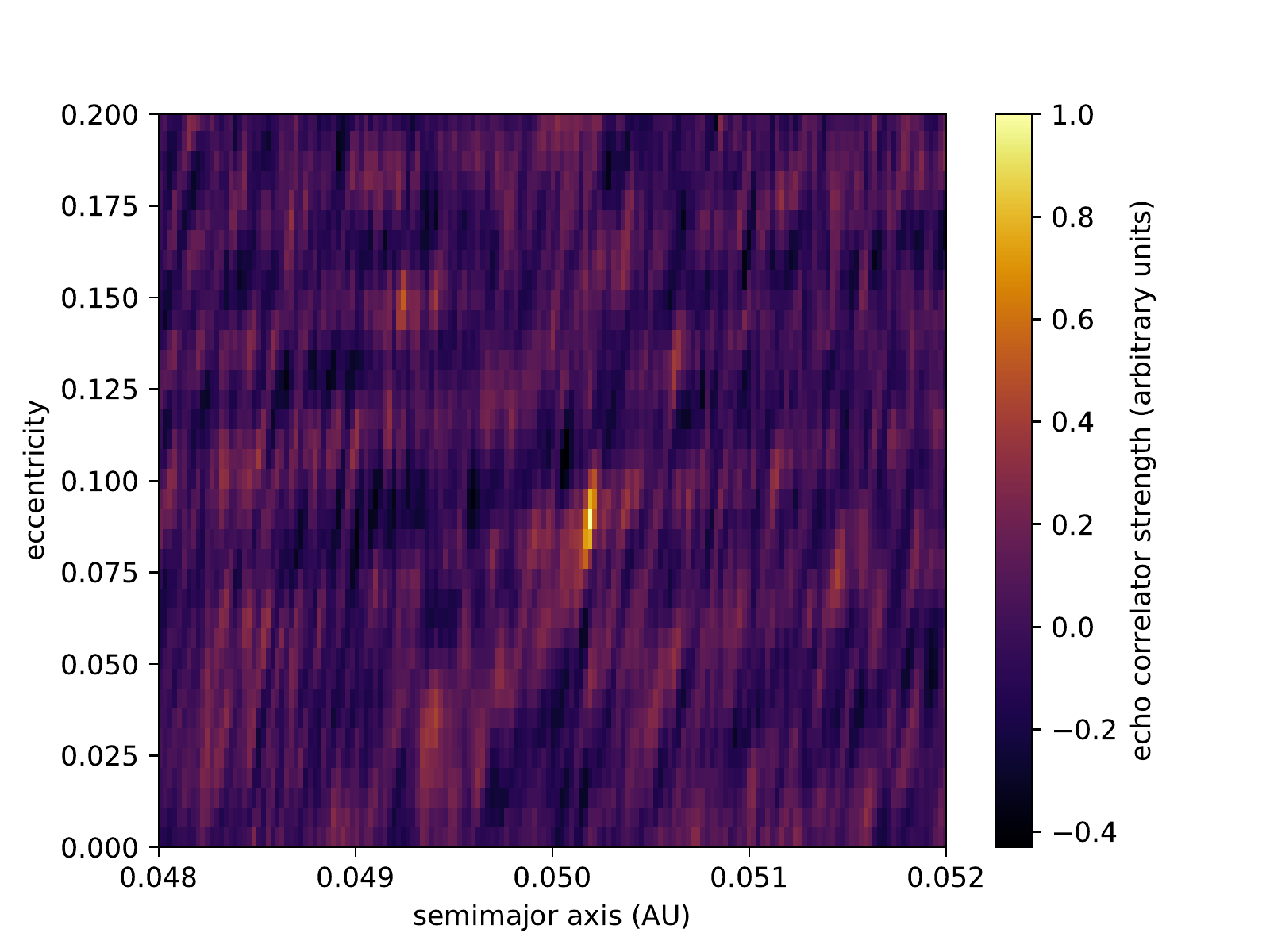}
\hspace{-0.25in}
\includegraphics[width=3.5in]{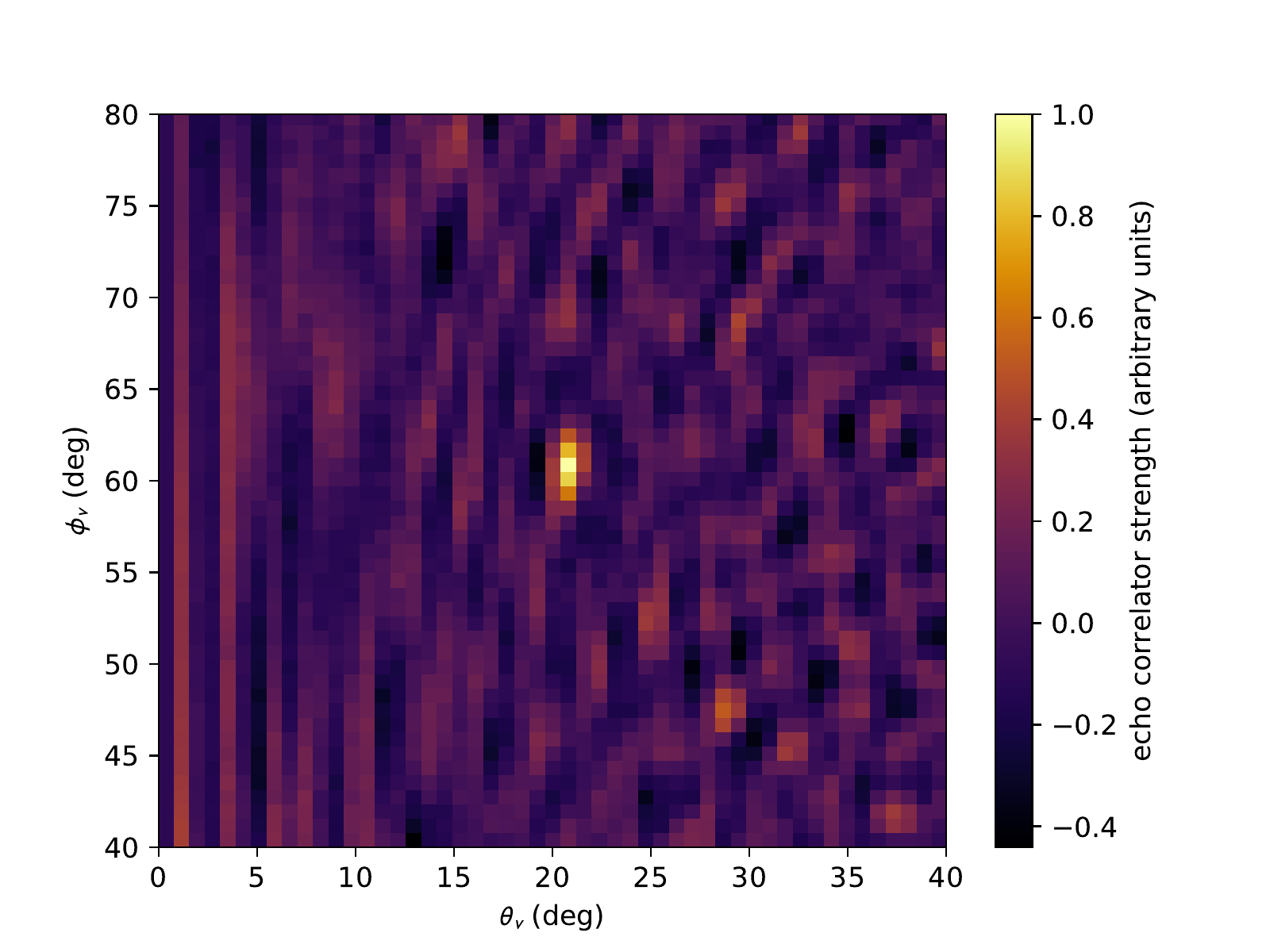}}
\caption{\label{fig:sixdim}Maps of the echo correlator strength from the set of simulated flares illustrated in Figure~\ref{fig:corr}, with echoes from a planet with orbital parameters $(a,e,M_0,\theta_v,\phi_v)$ set to $(0.05~\text{au},0.1,70^\circ, 20^\circ,60^\circ)$. 
The panels on are drawn from three billion unique samples of all five parameters and stellar mass (0.3~\Msolar). 
The peaks are near the true parameter values. 
}  
\end{figure}

In Figure~\ref{fig:sixdim}, we show maps of echo correlator strength in slices of the 6D parameter space ($M,a,e,M_0,\theta_v,\phi_v$) that contain the maximum value of $\mathcal{S}$. 
The peak value of $\mathcal{S}$ is close to its expected location; uncertainties derived from examining the region around the peak suggest that our method and choice of $\mathcal{S}$ is adequate for constraining the true orbital parameters.

This brute-force, 6D search method is an acknowledgment that finding the true orbital elements and observer orientation is a needle-in-a-haystack problem:
for the correct choice of parameters, and for a small region surrounding that point the parameter space, all the echoes line up in the estimation.
For all other choices of $\mathcal{S}$ they do not (see Figure \ref{fig:corr}).  
Thus a downhill search method will not be successful, except if a search is started very close to the true orbital parameters. 
In contrast, the dense, grid-based search demonstrated here will eventually find the true orbital parameters.

\begin{figure}[ht]
\centerline{\includegraphics[width=5in]{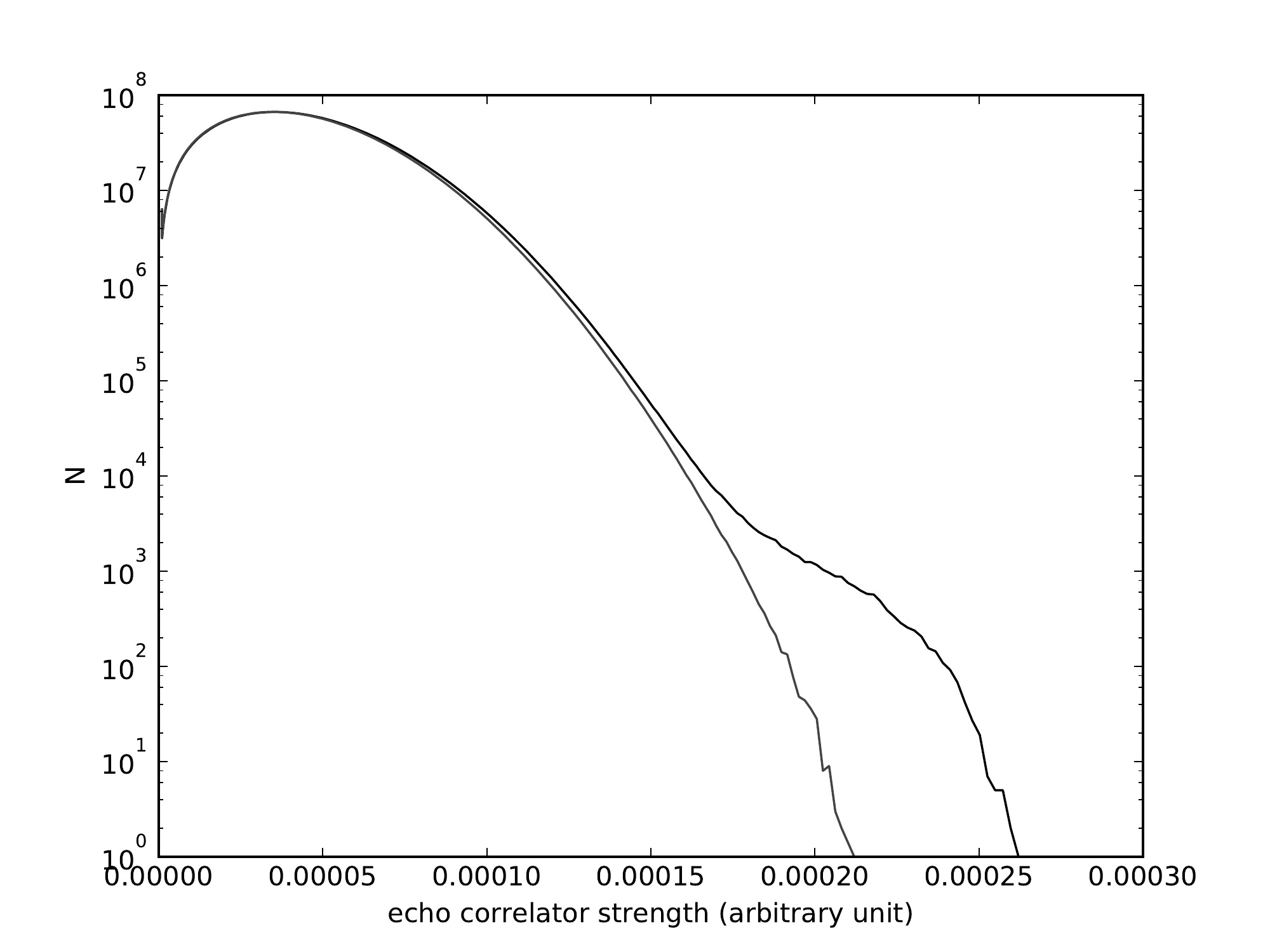}}
\caption{\label{fig:hist}Histograms of echo correlation strength measure $\mathcal{S}$. The dark histogram is from our set of 100 mock flares with echoes, while the light curve is from mock data with no echoes. 
The histograms compare well at low-$\mathcal{S}$ values, as in a lag-resampling analysis, and as reminiscent of the bimodal analysis in Fig.~\ref{fig:confidence}.}
\end{figure}

Initially, we may use the vast majority of points in our parameter search spaces that incorrectly align the echo delay time to derive noise limits in a ``hit-miss'' approach.
Because the vast majority of all fitting attempts are misaligned measurements, they sample a large number of different lag configurations.
Figure~\ref{fig:hist} compares $\mathcal{S}$ from mock noisy data in a broad 6D search, both with and without echoes. The comparison shows that the incorrectly aligned $\mathcal{S}$ values serve to sample the noise distribution, even though they are derived from data with echoes. 
This approach is not necessarily statistically unbiased because the samples here are correlated by orbital dynamics and their order of observation.
Nonetheless, Figure~\ref{fig:hist} suggests that, by omitting the region in the search space near a candidate echo signal, we can obtain a useful estimate of the one-point background distribution from which we may derive the statistical significance of a detection.

In our simulated data, the largest peak correctly corresponds to the signal produced by the exoplanet, and the histogram of all echo correlators (Figure~\ref{fig:hist}) shows a distinct knee in the distribution only when an echo is present.
However, we acknowledge that additional work will be required for echo detection to approach the rigor of existing detection methodologies.  
For example, in practice each flare will be unique, so some multi-peaked flares or significant post-flare brightening events may generate a significant correlation signal at a given time lag.  
While it may be tempting to lean heavily on a small number of extreme events that produce a strong signal, we caution that these outliers require special handling. 
To reduce experimenter bias, there are outlier rejection methods in the field of statistics that will likely play a role in any substantive discovery claim.
For example, the `jackknife' or leave-p-out methods are subsampling methods that leave out samples, then recompute the result to see how it changes--the calculation is repeated for each subsample permutation.
If a detection event is only present when a small set of the flares are included in the complete dataset, this would indicate that the effective sample size is much smaller than implied by the total number of flares collected and could call the result into question, depending on the specifics of the analysis.

One method we believe is promising for better identifying confidence intervals is lag resampling.
In both the 4D and 6D search models, excepting the face-on circular orbital case, the lag and echo contrast values change throughout the orbit.
Under the null hypothesis of no exoplanets, it should not matter if you reorder all of the flares.  
If there is an exoplanet, then reordering the flares will scramble the time lags and they will no longer add up coherently---the signal will be lost.
Therefore, by resampling the order of the flares with replacement (bootstrapping), it is possible to quantify how often a given detection magnitude occurs.
In performing this analysis, it is important to quantify the correlation of the resample: exchanging two flares that are co-added into the same lag bin will not resample the distribution.\footnote{There are some specific orientations and eccentricity combinations that can produce a single lag time, particularly when discretized to a single lag bin, comparable to face-on circular orbits.
Unlike face-on circular orbits, their echo contrast does not remain constant throughout their orbit.  
Therefore, if the detection algorithm being used includes a predicted orbital echo contrast weight, such as that described in Appendix~\ref{appx:weight}, it can still be fruitful to exchange two flares that would sum into the same lag bin.}
This lag resampling process can initially be performed just for the volume around the detection candidate.
However, in sampling an enormous 6D parameter space, there is always the concern of discovering a spurious correlation that might arise from data dredging.
Thus, to confirm any detection candidate, we suggest a full 6D bootstrap search of a resampled flare catalog to determine how likely false needles of the same magnitude would be discovered in the 6D haystack.
A definitive exoplanet echo claim will not only require a large amount of data, but also an effort from the astronomy community to develop and agree on rigorous detection criteria.
We hope that these flare models and simulation results motivate the beginning of that effort.

\needspace{6em}
\section{A framework for planet discovery}\label{sect:framework}

The feasibility studies performed in \S\ref{sect:method} indicate that it is challenging, though possible, to detect otherwise invisible exoplanets by their echoes. 
We therefore propose a generalized framework for the detection of planets from faint stellar flare echoes that involves a combination of signal processing and orbital estimation. 
The following steps provide an overview, while details of our demonstration implementation are provided in Appendix~\ref{appx:alg}.

\frameworkitem{Acquire data}
\label{sect:acquisition}
The most useful data will have a large flare signal, a high exoplanet albedo, and a low background signal. 
Also the cadence must be short, less than $\sim$10~s, by virtue of the short light-travel time for close-in planets.
While the majority of energy produced during a flare is often in the visible band, and stellar variability and exoplanet albedo are chromatic, the U-band appears to be most promising general choice because of the consistently small background signal, particularly for M dwarfs \citep[and references therein]{bromley1992, kowalski2016}.
Beyond photometry, it is known that flare emission from different lines and parts of the spectrum is not always synchronized: some spectral signatures may lag others and evolve differently on a timescale relevant to echo detection, which depends on the specific phenomenology of the star and the flare \citep[for example]{klocova2017}.
Running a cross-correlation on these signatures, often referred to as photoreverberation mapping, would then reveal a lag associated with the star's phenomenology, not with an echo.
On the other hand, if each spectral line can be linked to a known flare-exoplanet phenomenon, such as fluorescence, then it may be possible to use these signatures to greatly improve the detectability\citep{sparks2018}.
To exploit multispectral data using this manuscript's framework, for all reflected light spectral signatures, each band's light curve's autocorrelation signal will be synchronized in the lag domain, even if a given band has an arbitrary lag relative to the others.
Summing their autocorrelation signals can then further reduce the noise and uncertainty associated with a single flare event, though we have not explored the details of multiband autocorrelation in this study.

\frameworkitem{Process the light curve}
\label{sect:processing}
High-pass filtering, background subtraction, and detrending are essential to extraction of faint echoes in individual flares.
Higher-level processing of sets of flare light curves
can reveal additional information about active regions, such as star
spots, and reveal possible star-orientations, rotation rates, and even
differential rotation rates.  These parameters can be fed into models
to assist in estimating star mass, for instance.  Additionally, when
only a limited number of active regions are present, it becomes
plausible to predict which hemisphere around the star is illuminated
by a flare.  This additional information can be used as a `stellar
lock-in amplifier,' which could downweight flares that miss the exoplanet.
In general, this will knock down the noise much faster than histogram
analysis, but is only viable when the star is cooperative.

\frameworkitem{Generate flare catalogs}
\label{sect:catalog}
This step is to identify and select flare events and assign weights according to measures of quality, including flare strength and temporal structure or complexity.
Strong, delta-function flares are ideal, but broader, weaker flares can also produce echoes.  
In practice, each flare is assigned a weight; the flare strength may be a major component of the weight, but other factors can include a flare's width and structure, estimated by a frequency-domain analysis of the flare to identify the time-frequency bandwidth. 
We also recommend evaluating down-weighting any flares which exhibit detectable microflaring after the main pulse. 
While this may sound like a recipe for removing the exact data being sought, it is important to note that for most telescopes, particularly those under 10~m, the echoes will be near or below the noise level.
In practice, the astronomer can compute a threshold detectability for a given time lag using Eq.~\ref{eq:eps} and earlier work\citep{bromley1992} based on the specific observing conditions, and tag post-flare events above an appropriately optimistic threshold for potential masking.
This list is not exhaustive of weight possibilities; see the more detailed discussion in Appendix~\ref{appx:weight}.

\frameworkitem{Estimate autocorrelation functions}
The autocorrelation calculation is straightforward using discrete convolution routines.
However, it is often necessary to detrend the autocorrelation function.  
Because flares have a long tail, they can introduce curves to the autocorrelation, and the echo will be a small bump riding on its back.  
In simulations, we successfully reduced this tail in a variety of ways, including linear detrending, fitting a cubic polynomial to the logarithm of the curve, and linear Savizky-Golay filtering.
The optimal method typically depends on the flare structure.

\frameworkitem{Calculate echo correlator strength}
\label{sect:detection}
The autocorrelation curve signal strength will depend on the amplitude and detailed profile of the flare light curve.   
However, as the echo will essentially be an exact replica of the flare, its autocorrelation will also be a replica of the flare's autocorrelation.
Therefore, one means of calculating the echo strength is to extract the central peak of the correlator itself and convert it into a template for detection strength.
The detection strength parameter is maximized when it is customized for each flare based on that flare's structure, so the use of the autocorrelator itself is a self-adjusting, optimally `cusped' template.
Other templates include analytical filters such top-hat, triangle, or Savizky-Golay 2nd derivative filters---these are often tuned based on the width of the flare under analysis.
After producing the detection template, it is convolved with the correlator, resulting in a detection probability vector.
A discussion of this choice of echo correlator strength is given in Appendix~\ref{appx:alg}.

This process is performed for each flare, resulting in a detection matrix that can be readily indexed for the multi-dimensional search process.
As the values approach zero-lag, which corresponds to a time only moments after the flare, it is useful to de-weight the signal because it can be strongly influenced by edge effects.

\frameworkitem{Search for echo candidates}
\label{sect:search}
A first rapid search for echo candidates can use autocorrelation function estimates from the flare catalog to identify any high signal-to-noise events, or estimate any set of echoes that appear at constant lag.
Next, Lomb-Scargle periodogram analysis provides a low-overhead analysis of the data, to identify any potential high signal-to-noise candidates that are not necessarily at constant lag.
Beyond these glimpses into the data, it is necessary to align the detection strength parameters to the correct lag delay.
Therefore, we choose plausible ranges of stellar mass, a planet's orbital elements, and viewing parameters. For each set of values, we calculate time lags and extract echo correlator strengths from the flare catalog.
The resulting list of strengths is then summed; if this result exceeds a threshold, it is flagged as a potential hit for detection confidence analysis. An alternative approach can make use of an explicit model for a planet and its expected echo correlator strength, which is compared to data using  maximum likelihood analysis. This method may help eliminate false positives from microflaring.

We note that there are some orbital and viewing configurations that lead to degeneracies, at least in the predicted lag times.
For instance, when the argument of the periapsis is pointing away from Earth, then orbits with $e=\sin{\theta_v}$ can also have a constant lag, similar to a face-on circular orbit.
However, this degeneracy can be broken if we model the echo strength as a function of orbital phase. 
Here again, a maximum likelihood approach can help.

\frameworkitem{Determine detection confidence intervals}
\label{sect:confidence}
Two means of identifying true detection events based on confidence levels include hit-miss distribution analysis and lag resampling, as described above. The idea is to build up a model of the noise distribution, and compare candidates with this model. Our preliminary tests are encouraging.
We have also found that the structure of search results in the vicinity of a real detection event often has its own fingerprint.
This can be likened to the ambiguity function in radar signal processing, where the Doppler-shifted signal responds in a specific, predictable way.
Here, the slight misalignment of the echo lag parameters results in a predictable change associated with the specifics of the orbital parameters.
While we have not yet evaluated this echo lag ambiguity function in detail, it appears to be a promising additional route to perform detection analysis.

\needspace{6em}
\section{Prospects}\label{sect:prospects}

Ultimately, the astronomer seeking echoes will be at the mercy of each star's unique stellar activity.
But armed with a cooperative star, these results are meant to provide lower bounds and rough estimates of observing times and aperture sizes.
Using the model star parameters from \S\ref{sect:feasible}, we found an approximate lower bound of $\sim$75 flares for a star with $m_{U}=12$ and flare magnitude $\Delta m_{U}=-2$, observed by a 2-m diameter telescope with a 2~s integration time, to detect a planet at $a = 0.03$~au in a face-on circular orbit, with $R=2R_{J}$ and  $p=0.9$.  
However, with a 10-m diameter telescope using the same parameters, the signal-to-noise ratio is theoretically sufficient to detect an echo from a single clean flare.
In these cases, the challenge is to demonstrate that the signal is real and not an artifact of an additional microflare or post-flare brightening noise.
As with transit searches, multiple detection events are necessary for confirmation, and the various statistical tests described above are particularly helpful in this task.
To verify this, we ran the same bimodality test as \S\ref{sect:feasible} with 10~m telescope parameters and found that three hits and three misses result in a detection within the 98\% confidence interval under the bootstrap analysis.
We caution that extraordinarily clean flares and a high signal-to-noise are necessary for any claim based on an $N=6$, as well as a detailed analysis of the false alarm rate (the post-flare microflaring rate and characteristic observing noise from their instrumentation).

The framework for echo detection presented here depends on observing
campaigns that monitor flare stars over a sustained period of time
with as much collecting area as technically feasible. 
Even with 10~m-class telescopes, or perhaps arrays of smaller ones like MEarth
\citep{irwin2009}, only nearby stars are candidates. However,
prospects look good: Proxima Cen, with a known close-in planet
\citep{angladaescude2016}, is an excellent candidate for echo
detection; Trappist 1, a late-type M dwarf at 11~pc,
hosts seven planets \citep{gillon2017}. While it exhibits only
low-level flaring \citep{luger2017}, it illustrates that faint dwarfs
can have rich planetary architecture. Other active flare stars like AD
Leonis, EV Lacertae, YZ Canis Majoris, Wolf 359 (CN Leo), and UV Ceti
offer potential for new planet discovery.

To estimate the potential of echo detection, we consider the
2000 M dwarfs within a distance of 20~pc of the Sun 
\citep[Fig.~13, therein]{lepine2011}. Roughly half of nearby M dwarfs
show UV and X-ray emission from coronal activity \citep{stelzer2013},
similar to the fraction of M dwarfs in the Galactic plane inferred to 
have flare activity from the Sloan Digital Sky Survey \citep{west2004} 
and Kepler data \citep{walkowicz2011}.  In general, younger stars are 
more active \citep{giampapa1981,hilton2010}. The fraction of M dwarfs 
believed to harbor large planets with periods less that 
60 days is roughly 0.3 \citep{mann2012}; if we assume that the distribution 
of planets is constant with orbital distance, then about 10\%\ of these
stars have planets that are at least two Earth radii and inside of
0.1~au. Combining these estimates suggests that there may be roughly
100 candidate stars for echo detection within 20~pc. Perhaps a dozen
of these objects may be oriented to have transiting planets.

An effort to monitor these stars would be technologically difficult,
but the reward of detection is high. A discovery comes with full
orbital information about the planet, to within an angular rotation on
the plane of the sky. Narrowing down the orbital elements and
eccentricity can, for example, inform of planet formation processes
around low-mass stars \citep[e.g.][]{kennedy2006, kennedy2008a, kennedy2008b, ogihara2009, kb2014}. 
If fortuitous transits also occur, the planet's physical characteristics 
are also constrained. An echo detection may also contribute to the 
identification of multiplanet systems that are out of reach
of transit studies if the inclination distribution is broad.

To detect an exoplanet at 0.1~au, the relevant observing time scale is 50~s, depending on orientation.
It is necessary to have a gap in the light curve between the flare and echo, so there must be a minimum of one data point in between, pushing the minimum cadence down to around 25~s.
The autocorrelation signature is further enhanced if it resolves the fine structure of the flare, so useful observation timescales are nearly all faster than a 10~s cadence.
Fortunately, several ongoing and upcoming missions are starting to regularly provide the necessary collection rates for observation.
The TESS spacecraft captures images every 2~s (though the data products are summed into 2~minute intervals for downlink, and the collection area is only a few inches in diameter)\citep{tess2017}.
The Transneptunian Automated Occultation Survey (TAOS II) will have a readout cadence of 20~Hz and over a meter collection area, making it a promising platform for a preliminary survey\citep{taosii2012}.
The planned PLATO spacecraft would provide potential detection capabilities with its 2.5~s cadence and long dwell times\citep{plato2016}.
The proposed LUVOIR spacecraft would provide exceptional detection capabilities, depending on the final embodiment, with a proposed 8-15~m telescope and dedicated ultraviolet photometric bandpass sensors\citep{luvoir2017}.
The proposed HABEX spacecraft would also provide excellent detection capabilities, depending on the final embodiment\citep{habex2016}.

\needspace{6em}
\section{Conclusion}\label{sect:conclude}

Here we introduce a framework for extracting and interpreting faint planetary echoes of stellar flares.
It opens up a new detection regime for planet hunting, namely the very inner region of any planetary system, not just those that happen to be viewed edge on. 
This method complements existing detection techniques well, and is feasible with existing instrumentation. 
Many observed flare-like events are required for detection, but by prioritizing active stars, we can reduce the observation time needed to demonstrate the feasibility.
Additionally, because our approach only requires photometry, multiple stars can be monitored simultaneously with a single focal plane array---an echo detection survey can piggyback on the hardware of transit and asteroseismology surveys if they provide support for $\lesssim$10~s cadence data products.

With simple simulations, we also show that it is possible for echoes to reveal information about a planet and its stellar host.
In our examples we cover orbital orientations that are close to the host star, viewed face-on or approximately so, to exemplify the strength of this method for discovering new planets in a previously undetectable regime of orbits. 
Our statistical echo detection method is effective for all bright close-in planets, whether observed face-on, edge-on or somewhere in between.
Given enough events, we are able to extract the complete set of orbital parameters for an exoplanet.

The main difficulty with our method is that the echoes are drowned out by noise in most situations. 
To maximize the strength of an echo against the noise of a variability event and prevent scintillation-induced false positives, space observation is strongly beneficial. 
Another host of difficulties may come from complex flare structures that can mimic or mask echoes. 
Other complications come from non-idealities of the flares, including the change in light travel time associated with where they the occur on the star's surface and their potential spatial extent.
Ways to obviate this problem include downweighting, rejecting flares with multiple large peaks, more detailed detection algorithms that can accommodate more complex parameters such as a maximum likelihood model, or developing different detection algorithms based on deconvolution. 
Future work will consider this problem using observed flare light curves.

While echo detection requires large data sets and extensive multi-parameter searches to produce convincing results, we are emboldened by the success of Kepler and other persistent observatories that provide years of data.  
Furthermore, the capacity to extract orbital parameters provides a major benefit to other exoplanet endeavors, including studies of habitability and validating solar system formation models.  
And echoes rely on different phenomena from other planet hunting techniques, providing the opportunity for an independent confirmation of a discovery.
In light of the unique advantages of echo detection, we believe that the challenges are worth overcoming.

\acknowledgements 

This research has made use of the Exoplanet Orbit Database and the Exoplanet Data Explorer at exoplanets.org.  
This research is supported by NASA Grant 80NSSC18K0041 and NNX17AE24G.  
We acknowledge generous allotments of computer time on the NASA `discover' cluster.
We thank the referees for the careful reading of our manuscript and the comments, which resulted in significant improvements to the paper.

\appendix

\needspace{6em}
\section{Autocorrelation and Noise Analysis}\label{appx:autocorr}

We define the autocorrelation function of the light curve as
\begin{equation}
 C(\tau) = (L * L)(\tau) \equiv \frac{1}{T_{obs}} 
\int_{t_i}^{t_i+T_{obs}} L(t^\prime)L(t^\prime-\tau) dt^\prime
\end{equation}
where $t_i$ is the starting time of the observation and $T_{obs}$ is the duration.  
In practice, we use a discretized version of this integral expression, and only after detrending the light curve, which removes any DC component \citep[e.g.,][]{argyle1974}.  
In terms of the flare, echo and noise contributions, after subtracting the constant background signal ($Q$) from Eq. \ref{eq:lightcurve},
\begin{equation}\label{eq:C}
\begin{split}
C(\tau) &=  (L*L)(\tau) \\
&=  (F*F)(\tau) + \epsilon (F*F)(\tau-\techo) + \epsilon (F*F)(\tau+\techo) + (F*N)(\tau)+ (F*N)(-\tau) \\
&+ \epsilon(F*N)(\tau- \techo)+ \epsilon(F*N)(\tau+ \techo)+\epsilon^2(F*F)(\tau) + (N*N)(\tau). 
\end{split}
\end{equation}

We can simplify Equation~(\ref{eq:C}) by removing the echo-echo and echo-noise convolution terms when the noise is of the same scale as the echo, both of which go as $\mathcal{O}(\epsilon^2)$.
Then, keeping leading order in the noise and in the echo contribution, we find
\begin{equation}\label{eq:Capprox}
C(\tau) \approx (F*F)(\tau) + \epsilon (F*F)(\tau-\techo)+ \epsilon (F*F)(\tau+\techo)  + (F*N)(\tau)+ (F*N)(-\tau) + (N*N)(\tau).
\end{equation}
For Gaussian white noise, for finite data of length $n$, the noise-noise term produces background on the order of $StdDev((N*N)(\tau)) \approx \sqrt{n-\tau} Var(N(t))$, so it is important not to use an excessively large window.
To accommodate the flare-flare term, $(F*F)(\tau)$, we typically detrend the autocorrelation function.  
This assumes that the flare tail is smooth, which may not always be the case.
The remaining challenge is to identify an echo signal, $\epsilon (F*F)(\tau-\techo)$, which is buried in convolved noise, $(F*N)(\tau)+(F*N)(-\tau)+(N*N)(\tau)$.  
To produce a preliminary estimate, we treat the flare term as a delta function, giving:
\begin{equation}\label{eq:cReduced}
C(\tau) \approx \intflarepeak^2 \delta(\tau) + \intflarepeak \epsilon \delta(\tau-\techo)+ \intflarepeak \epsilon \delta(\tau+\techo)+ \intflarepeak N(\tau)+ \intflarepeak N(-\tau) + (N*N)(\tau),
\end{equation}
where $\intflarepeak$ is associated with the flux from the flare, independent of the stellar background.
At the time of the echo, the signal is:
\begin{equation}\label{eq:catecho}
C(\techo) \approx \intflarepeak \epsilon + \intflarepeak N(\techo)+ \intflarepeak N(-\techo) + (N*N)(\techo).
\end{equation}
The goal is to identify a criterion for when the echo, $\intflarepeak \epsilon$, is distinguishable from the autocorrelation background noise.

The noise terms at $\pm \techo$ are uncorrelated, so they should add in quadrature.
The residual term is the convolved noise-to-flare-peak ratio, which is only slightly biased by the noise terms evaluated at $\pm \techo$, so we treat them as independent.  
With these assumptions, we can treat this as a signal detection problem of two normal distributions of equal standard deviations, $\sigma_d=\sqrt{2(\intflarepeak \sigma_{N})^2 + (n-\tau) \sigma_{N}^4}$, separated by sensitivity index $d'=\intflarepeak \epsilon/\sigma_d$.
For counting noise, $\sigma_{N}=\sqrt{Q\deltat}$, giving:
\begin{equation}\label{eq:detectionsensitivity}
d'=\frac{\intflarepeak \epsilon}{\sqrt{2\intflarepeak^2 Q\deltat + (n-\tau) (Q\deltat)^2}}
\end{equation}
In the large flare limit, $d'=\epsilon/\sqrt(2Q\deltat)$.

\needspace{6em}
\section{Lags and Echo Contrast for Different Orbits}\label{appx:orbitalsignals}

\begin{figure}
	\centering
	{\includegraphics[width=1\textwidth]{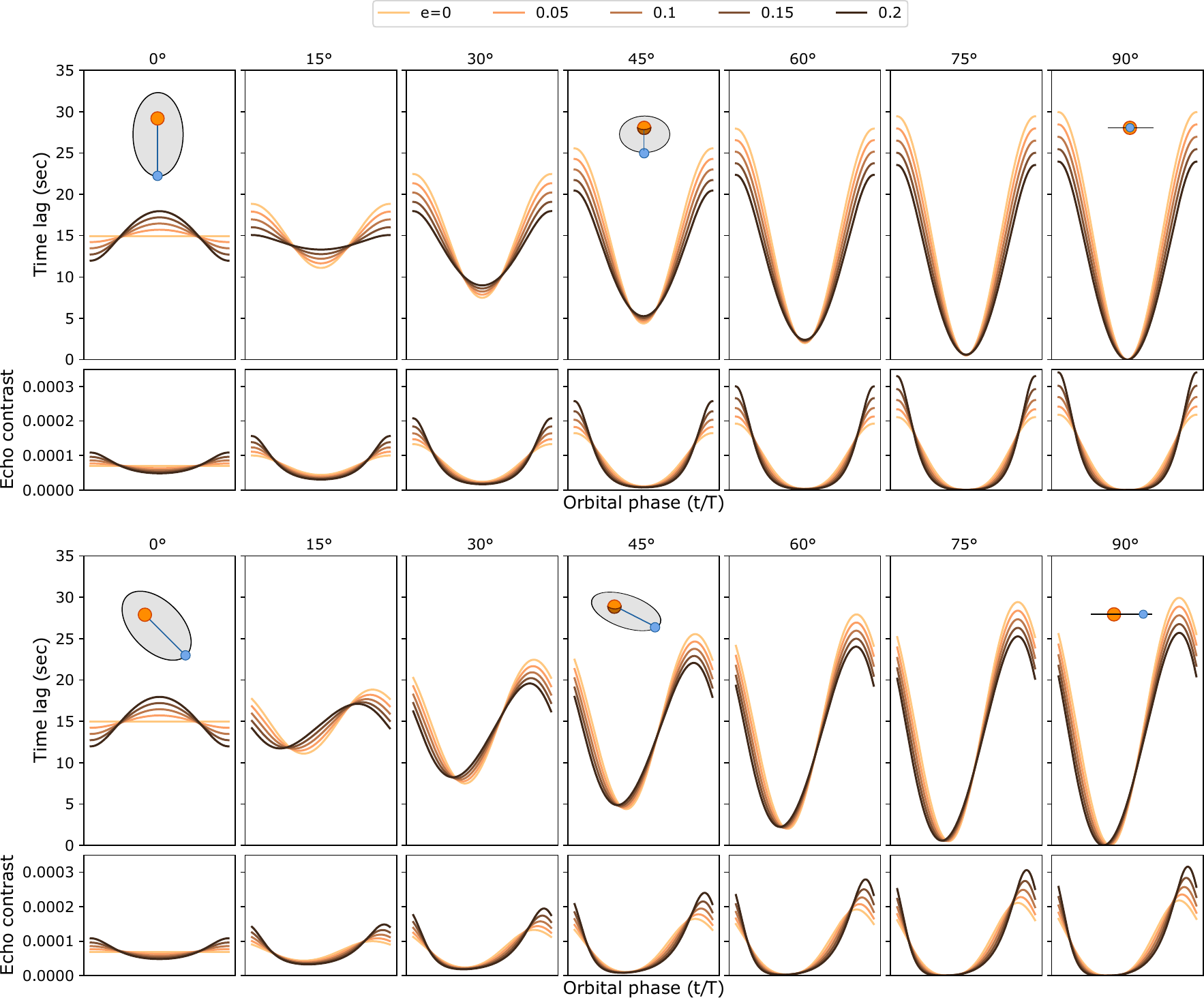}}
\caption{\label{fig:peri0}Jupiter-sized exoplanet at 0.03~au at eccentricities from 0-0.2 with argument of the periapsis oriented at 180$^o$ (Top) and $135^o$ (Bottom) relative to Earth.}
\end{figure}

\begin{figure}
	\centering
	{\includegraphics[width=1\textwidth]{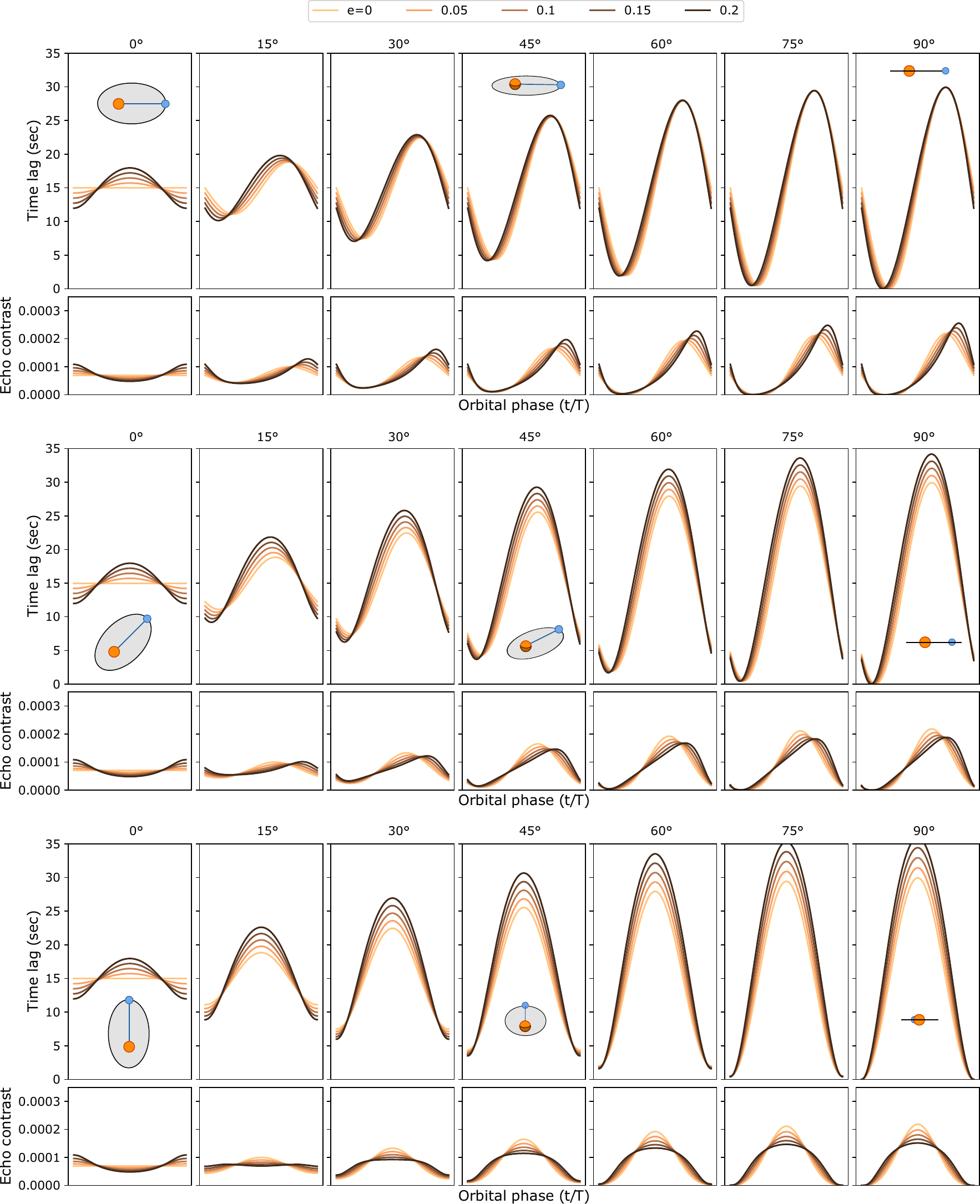}}
\caption{\label{fig:peri90}Jupiter-sized exoplanet at 0.03~au at eccentricities from 0-0.2 with argument of the periapsis oriented at 90$^o$ (Top), $45^o$ (Center), and $0^o$ (Bottom) relative to Earth.}
\end{figure}

Each orbit and orientation relative to Earth will have different lag and contrast as functions of orbit.
To predict the observability for different orbits, we calculated the lag and planet phase function as a function of different orbital parameters using Eq.~\ref{eq:echoorbit} and a Lambertian phase function \citep{traub2010}:
\begin{equation}\label{eq:lambertian}
\phi(\alpha)=\frac{\sin{\alpha}+(\pi-\alpha)\cos{\alpha}}{\pi}
\end{equation}
with $\alpha$ the orbital phase of the exoplanet.
Figs.~\ref{fig:peri0} and \ref{fig:peri90} show examples of numerically computed lags and contrast for different eccentricities and viewing angles.
These are similar to the delays discussed in Mann\citep{mann2017} and Sparks\citep{sparks2018}.

\begin{figure}
\centerline{\includegraphics[width=5.0in]{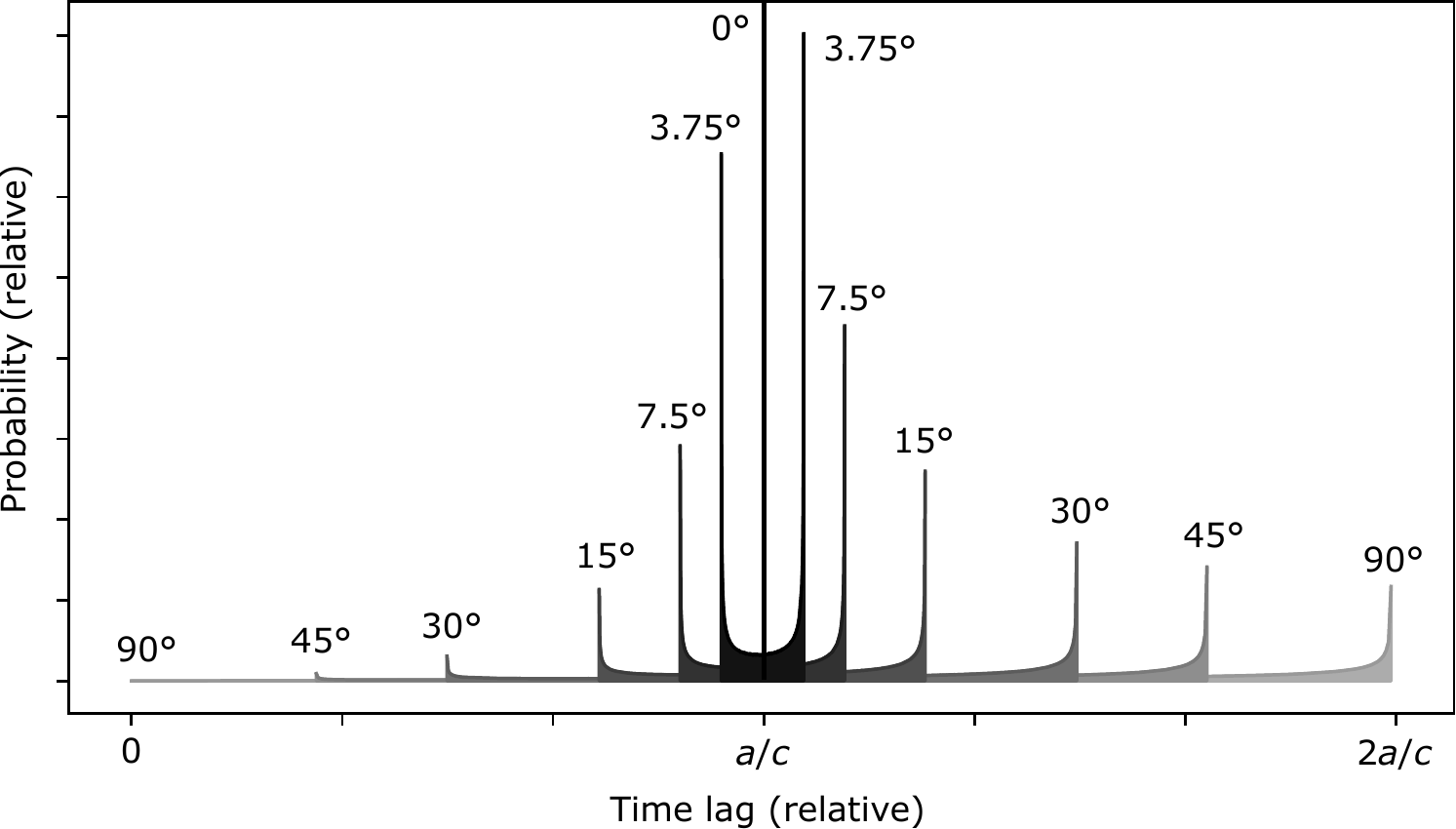}}
\caption{\label{fig:circprob}Probability of an exoplanet in a circular orbit being visible to Earth for a given time lag as a function of inclination angle, weighted by flare view factor and assuming a Lambertian exoplanet.  
For face-on orientations, the probability function collapses to a single point at the center. 
For the transit case, the exoplanet is most visible when it is in superior conjunction.}
\end{figure}

While this can be evaluated numerically for all orbits, we found that the probability of detecting echoes from an exoplanet in a circular orbit observed from an inclined view has a simple analytical form.
To see this, we start with the lag as a function of time\citep{sparks2018}:
\begin{equation}\label{eq:lagt}
    \techo(t) = \frac{a}{c}(1+\{\sin{2\pi t/T}, \cos{2\pi t/T}, 0\}.\hat{e}_\text{obs}) = 
		\frac{a}{c} (1+\sin{2\pi t/T} \sin{\theta_v})
\end{equation}
This is readily inverted to give:
\begin{equation}
	t(\techo) = \frac{T}{2 \pi} \arcsin{\frac{c \techo}{a}-1/\sin{\theta_v}}
\end{equation}
This form is only valid where arcsin is defined, which provides us with maximum and minimum lags of
\begin{equation}
\frac{a}{c} (1-\sin{\theta_v}) < \techo < \frac{a}{c} (1+\sin{\theta_v})
\end{equation}
This can be normalized to the arcsine cumulative density function, which has a variety of well-studied properties.
One of the most important properties is that the modes of the distribution occur on the ends, and not near the mean (the arcsin probability density function is of the form $P(x)=1/\pi \sqrt{x(1-x)}$, which is infinite at $x=0, 1$).  
Namely, there are two stationary points where the exoplanet would be more readily observed, shown in Figure~\ref{fig:circprob}.  
In the plot, we have also included the phase function weights.

\begin{figure}
\centerline{\includegraphics[width=.7\textwidth]{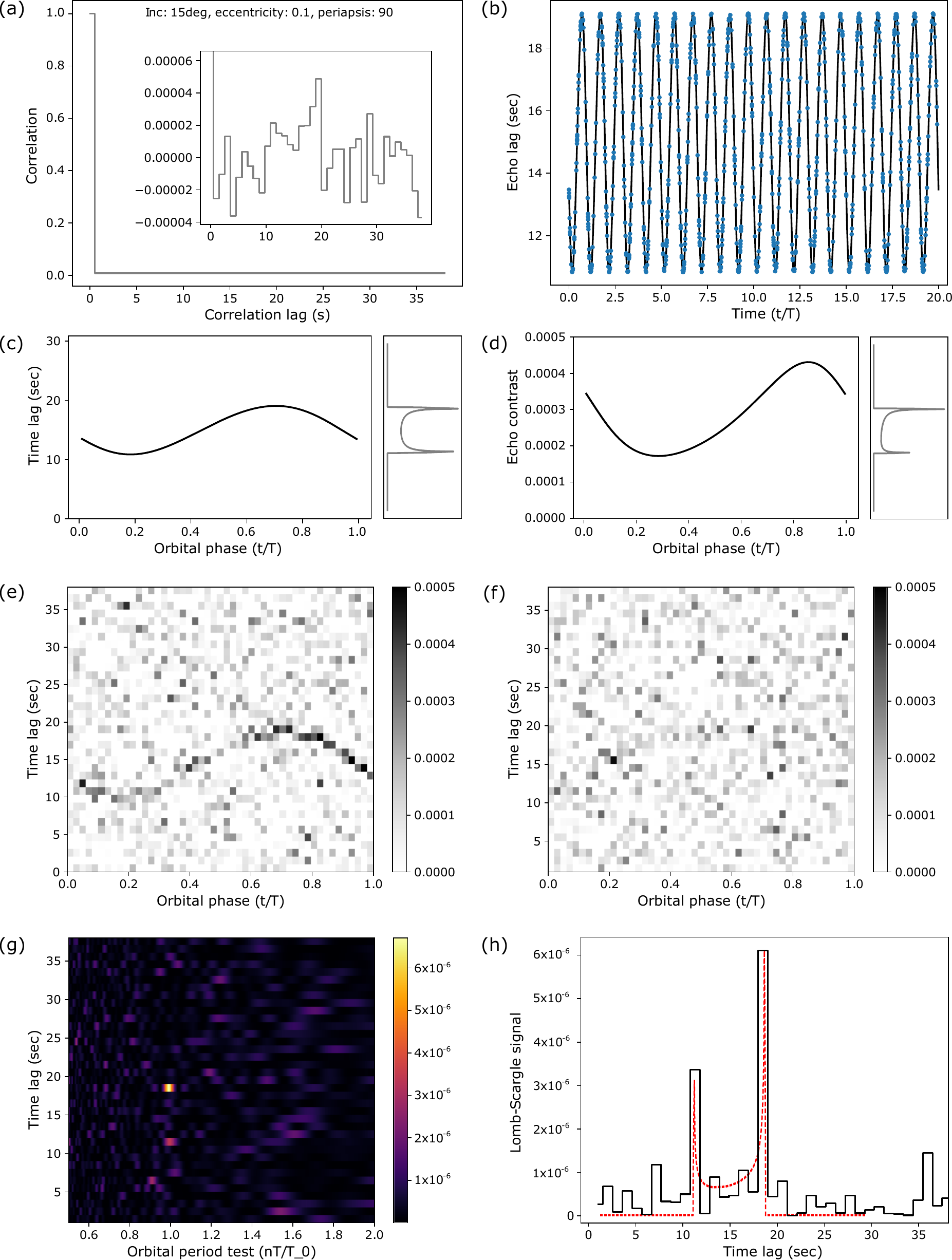}}
\caption{\label{fig:periodicity}
Representative results for 1000 flares occurring over 20 periods for an exoplanet at 0.03~au, $R=2R_J$, orbital inclination of 15$^o$ relative to Earth, eccentricity of 0.1, and argument of the periapsis at 90$^o$ relative to Earth. 
(a) The simulated autocorrelation is not pristine due to the presence of multiple lags. 
(b) Analytical values for the echo lag as a function of orbit, as well as the random times that flares were sampled. 
(c) Phase-folded lags and their associated histogram (right) showing the most probable lags. 
(d) Exoplanet-star contrast as a function of orbit, used as a surrogate for echo contrast, and the impact on the lag detection histogram. 
(e) The orbital phase folded autocorrelation result when the exact phase is used.
(f) The orbital phase folded autocorrelation result when the period is wrong by 10\%.
(g) Lomb Scargle plot for the detection showing two bright peaks associated with the orbital position.
(h) Cross-section of the Lomb Scargle plot at the correct period, showing the same peaks in relatively similar structure as the predicted form from (c) and (d).}
\end{figure}

Given that many of these lag structures will produce characteristic peaks, it is plausible to search for them without performing complete multi-dimensional searches.
To evaluate the feasibility of performing lower-dimensional searches, we produced a simulation for a hot gas giant with of $R=2R_J$ around a star with numerous delta-like flares.
Figure~\ref{fig:periodicity} shows data analysis results for 20 simulated periods for  orbiting an active flare star.  
Fig.~\ref{fig:periodicity}(a) shows the 1D correlation function, which has some promising peaks, but they are not sufficiently structured to be definitive.
Fig.~\ref{fig:periodicity}(b) shows how the flares and echo lags are distributed, illustrating why a 1D analysis is inadequate.
Fig.~\ref{fig:periodicity}(c) shows the ideal time lag and the associated distribution of lags as observed over a single period.  
Two prominent peaks are present with a higher density between them.  
A more realistic distribution is shown in Fig.~\ref{fig:periodicity}(d), where the peaks have been weighted by the ideal exoplanet-star contrast. 
Indeed, this double-peak structure could be used as a matched filter search of the 1D lag function.
Fig.~\ref{fig:periodicity}(e) shows a period-folding analysis, showing higher signal strength in the predicted regions if the period is known precisely, but if the folding period is slightly off, say by 10\% (Fig.~\ref{fig:periodicity}(f)), the signal disappears.
A comprehensive period-fold search is similar to the 4D approach described in \S\ref{sect:method}.
Alternatively, the workhorse of periodic searches is the Lomb-Scargle plot.
The same dataset processed in this way produces Fig.~\ref{fig:periodicity}(g), which shows a single significant peak.
Examining only the period with this peak, Fig.~\ref{fig:periodicity}(h), the structure of the two peaks is apparent and closely matches the distribution in (d), overlaid as the red dotted line.  
Again, the double-peak structure could be used as a matched filter search of a Lomb-Scargle plots.

\needspace{6em}
\section{An Echo Detection Algorithm}\label{appx:alg}

\begin{enumerate}

\item Produce a flare catalog.  We identify flares in light curves using a
  trigger based on relative change in flux, and extract short segments
  of light curves that span individual flares and potential echoes. In
  the tests with simulated data that we perform here, each segment
  contains the integrated flux in $N_t\sim 10^3$ bins of width $\Delta
  t = 0.1$~s, and covers a time $t_\text{seg}$ of 100~s, with the
  flare offset from the start of the segment by $\tau_f$, the typical
  flare decay time scale.  We remove slowly varying modes in each
  segment by applying a high-pass filter,
  \begin{equation}
    H(f) = 1 - \exp(-f^2/2\sigma^2)
  \end{equation}
  where $f$ is frequency, and $\sigma = 2\pi/t_\text{cut}$, such that
  cut-off time $t_\text{cut}$ is three times $\tau_f$.  The filtering
  is performed with fast Fourier transforms using the Python NumPy package\citep{numpy2006}.

\item Generate the discrete autocorrelation function, or ``correlator,'' of each light curve segment. 
	The correlator for the $i^{\text{th}}$ light curve segment is
  \begin{equation}\label{eq:xi}
    \xi_j^{(i)} = \sum_{k=0}^{N_t-j} \tilde{F}^{(i)}_k \tilde{F}^{(i)}_{k+j}
  \end{equation}
  where subscript $j$ designates the correlator's lag, $\tau = k\Delta t$, $\tilde{F}^{(i)}_k$ is the filtered flux in a time bin at $t = k\Delta$ (relative to the start time of the light curve segment, $T_i$). 
	To calculate the estimator in equation~(\ref{eq:xi}), we use the Python NumPy correlate routine in ``full'' mode.\citep{numpy2006}
	Then, to help with echo signal extraction, we flatten the correlator by subtracting a cubic fit to $(\log(t),\xi)$ on a domain with $t_\text{cubic} < t < t_\text{seg}$ (NumPy polyfit), where $t_\text{cubic}$ is chosen to be the typical flare decay time, $\tau_f$\citep{numpy2006}.

We define a template function to identify echo candidates. 
 Given that the shape of the echo in the correlation function should 
 closely match the shape of the correlator at zero lag, we 
 typically use a truncated, scaled version of the correlator itself.

\item Define a template function to identify echo candidates. 
 Given that the shape of the echo in the correlation function should 
 closely match the shape of the correlator at zero lag, we 
 typically use a truncated, scaled version of the correlator itself,
  \begin{equation}
    {\Xi}^{(i)}_j = {\xi}^{(i)}_j \ \ \ \text{($|j| \leq J_0$;
      zero otherwise)},
  \end{equation}
  where $J_0$ correspond to a lag that is four time the value at which
  the average autocorrelation function drops to half of $\bar{\xi}_0$.
  This choice sets the half-width of the template to be approximately
  the flare decay time. We further set the average value of each
  template to zero to avoid coupling with the local correlator
  background near any potential echo location.

\item Determine a weight, $W_i$, for each of the $N_f$ correlators to
  provide a measure of the quality of each flare. The noise level is
  important, as is the orbital phase, but so is the flare light curve's
  profile. For example, a flare that has all photons launched in a
  single time bin (a delta-function flare) might be high quality,
  while a brighter flare that is spread out over a long period of time, longer
  than the echo delay times of interest, could have lower quality. 
  See Appendix \ref{appx:weight} for a further discussion of weighting.

\item Set parameters needed to predict the echo delay for a planet in
  orbit about the flare star. As in the main text, we choose a 6D
  vector $\vec{P}$, with elements $M$, $a$, $e$, $M_0$,
  $\theta_v$ and $\phi_v$, where $M$ is the stellar mass, $a$ is the
  semimajor axis, $e$ is the orbital eccentricity, $M_0$ is the mean
  anomaly at some reference time $T_0$, while the last two parameters
  are spherical polar angles that determine the direction to the
  observer. We orient a coordinate system so that the planet lies in
  the $x$-$y$ plane ($\theta_v = \pi/2$) with periastron on the
  positive $x$-axis ($\theta_v = \pi/2$, $\phi_v = 0$), and the
  planet's orbital angular momentum vector defines the
  $z$-axis. Thus when a planet is observed face-on, $\theta_v =
  0$. 
	
	For now we assume that all flares are centered on the host
  star, and discuss below how to relax this condition. Note that in
  general we may not always know the stellar mass accurately, and so
  may have to include the stellar mass $M$ as a sixth ``orbital''
  parameter.  Here, we assume $M$ is known.

  For a given set of orbital parameters $\vec{P}$, we shift and add
  the weighted correlators so that echo signals align and reinforce
  each other. The shift is calculated from the echo delay time of each
  flare,
  \begin{equation}
    J_e = (\techo - a_f/c)/\Delta t_\text{bin}
  \end{equation}
  so that echo signals are centered in bin $\bar{J}_e$, corresponding
  to a lag of $a_f/c$, as if the planet were on a circular orbit at
  some fiducial semimajor axis $a_f$, observed face-on.

\item Build up the fingerprint of an echo. We start
with a shifted, weighted correlator,
  \begin{equation}\label{eq:echoesignal}
    S^{(i)}_j(\vec{P}) \equiv 
     W_i \, g_{j-J_e} \, \xi^{(i)}_{j-J_e},
  \end{equation}
  where $W_i$ is the flare quality weight, and the function $g_{j-J_e}$
  is unity so long as the lag shift does not move the peak of the
  correlator (originally at zero lag) close to the 
  mean echo delay lag.\footnote{This restriction limits how close to
    the host star a planet can be. It also means that for an edge-on
    orbit, we will be ignoring the echo events near transit, which
    also have very small echo delays. However, in this case the echoes
    themselves will be weak, since the planet will be in a crescent
    phase.}  In practice we prevent the shifting process from
    moving the flare peak to within twice the typical flare decay time
    of the mean echo delay time, though carefully chosen asymmetric 
    templates may be able to provide echo correlation estimates.  
    We note that the shortest echo lags typically correspond to orbital 
    phases that are already unfavorable for detecting reflected light, 
    so these data points are generally of lower detection value. Then 
  \begin{equation}\label{eq:echoesstrength}
    \mathcal{S}(\vec{P}) \equiv \frac{1}{W} \sum_{i=1}^{N_f}
                  \sum_{j=-J_0}^{J_0}
                  \Xi_j \, S_{\bar{J}_e-j}.
  \end{equation}
with $W$ the sum of weights.  
The quantity $\mathcal{S}(\vec{P})$ is an ``echo correlator strength'', which
serves as a measure of success in a search for echoes in a set of
flares.

Alternatively, with the same data, one form of a likelihood signature is (though it would typically be summed in log form):
  \begin{equation}\label{eq:maxlikely}
    L(\vec{\xi}|\vec{P}) = \prod_{i=1}^{N_f}
									\frac{1}{\sqrt{2\pi \sigma(W_i(j))^2}} \exp{\frac{-[\breve{\epsilon}^{(i)}_j(\vec{P})-(\Xi*S)(j)]^2}{2 \sigma(W_i(j))^2}}
  \end{equation}
where $\breve{\epsilon}$ is the predicted echo correlator strength based on $\vec{P}$ for the observed flare, $j$ is the lag anticipated by the model at flare time $t$, and $(\Xi*S)(j)$ is the template convolution term from Eq.~\ref{eq:echoesstrength}.  
To be strictly accurate, this requires making assumptions about the exoplanet radius and albedo; $R=R_J$, $p\sim0.7$ is relatively robust first guess for the majority of echo-observable exoplanets.
This form is also amenable to more detailed models, such as multi-planet systems, and is less likely to be contaminated by including a large number of microflares.
The structure of $\sigma(W_i(j))^2$ is briefly discussed in Appx.~\ref{appx:weight}.
This signature would then be maximized by a search routine.

\item Map the echo correlator strength over orbital elements, viewing
  angles, and (if necessary) stellar mass. We densely sample the 6D
  orbital parameter space on a grid, measuring $\cal{S}$ using
  equation~(\ref{eq:echoesignal}) at each grid point, $\vec{P}$.  This
  procedure may involve as many as $10^{10}$ points, needed to map out
  $O(100)$ parameter values in typical ranges: $0.01~\text{au} < a <
  0.1~\text{au}$, $0<e<0.5$, $0^\circ<M<360^\circ$,
  $-90^\circ\theta_v<90^\circ$ and $0^\circ<\phi_v<180^\circ$).  Note
  that there is a degeneracy in viewing angle values, since echo
  delays are symmetric to $(\theta_v, \phi_v) \rightarrow
  (-\theta_v,\phi_v+ \pi)$, so we only need to cover half the possible
  range of one or the other viewing angle. We find that we need the
  finest grid for the semimajor axis, and recommend a grid spacing $\Delta
  a$ that does not exceed
  \begin{equation}
    \frac{\Delta a}{a} \sim 0.01 \frac{T_{\text{period}}}{T_{\text{total}}}
  \end{equation}
  where $T_{\text{period}}$ is the orbital period an
  $T_{\text{total}}$ is the total span of the observations. The reason
  is that a more accurate value of the period (hence $a$) is needed to
  line up the echo signals if the observations cover a larger number
  of orbits. 
	This has the effect of requiring a smaller $\Delta a$ for longer observation times.
	Our simulations have only considered upwards of 10's of periods, but for multi-year observations, such as those performed by Kepler, $\Delta a$ could become very small.\footnote{This problem has been encountered in other fields, such as radial velocity studies. 
	One solution is to use Welch windows\citep{welch1967} --- select a maximum coherence length, break the data into overlapping windowed segments of that length, compute the periodograms on each segment, then sum them.}

\end{enumerate}

\needspace{6em}
\section{Flare Weighting}
\label{appx:weight}

This appendix is included as a discussion of potential weighting methods to increase the effectiveness of detection algorithms and is provided to motivate future work.
To account for the fact that not all flares are equally useful, it can be helpful to weight the flares.
The flare-background magnitude is the most obvious constraint. If the flare is 1\% of the background signal, 
and the echo is $10^4$ times smaller, you will need more than a few measurements. Assuming Poisson noise, 
the average integrated noise for the quiescent background $Q$ over period $t_{int}$ is:

\begin{equation}
\langle N \rangle = \sqrt{Q t_{int}}
\end{equation}

For a flare with an integrated number of counts above the baseline of $f$, with length $t_{flare}$ the flare S/N is roughly:

\begin{equation}
s = \frac{f}{\sqrt{Q t_{flare}}}
\end{equation}

Therefore, a straightforward weight for the autocorrelation function
of flare event $i$ is $s_i$. In this way, a small, brief flare can
often have the same weight as a larger, long flare. However if the
flare is bright and broad and not particularly useful, we look at the
quality of the flare.

For a given orbit, the exoplanet will have a reduced probability of being viewed in certain positions, in accordance with its phase function.
In simulating a candidate orbit, the predicted echo magnitude can be used as a weight in order to accentuate the highly visible parts of the orbit and to suppress orbital phases for which the echoes are fainter and prone to noise.
Typically, a Lambertian approximation is used.
Appendix~\ref{appx:orbitalsignals} shows several examples of echo magnitudes, these would serve as the phase weight.

We can determine a \textit{quality} for any given fluctuation from its
time-bandwidth product properties that can be used as a weight. Using the time-frequency
uncertainty relation ($\thalf \Delta \omega \ge \frac{1}{2}$) and the
threshold property found by localizing an echo to within half the
travel time from the star to exoplanet of $\thalf \gtrsim \techo/2$, we
can see the spectral bandwidth of a flare's light curve must be at least

\begin{equation}
\Delta \omega \gtrsim \frac{1}{\techo}
\end{equation}

Variability events with small spectral widths can be culled (a sine
wave is useless, which has a delta-function spectral width). A good
metric for the width of a pulse, $\Delta \omega$, is the inverse of
the magnitude of the second moment of the Fourier transform of the
flare. Thus, to weight the contribution to the autocorrelation
function we can use $w_i = (\Delta \omega )_i$ for variability event
$i$.

Not all fluctuations detected from Earth will occur on the same side of the star as the exoplanet, resulting in a systematic decorrelation if the data are blindly summed. 
For a face-on orbit, it is reasonable to expect about half of the localized fluctuation events (e.g. flares) to occur on the same side of the star as the exoplanet, and many that occur near the limb will be attenuated by the optical density of the star, depending on the depth of the fluctuation. 
For nearly edge-on systems, the odds become worse because the exoplanet's phase function will be at its brightest at superior conjunction, which means it will primarily be illuminated by events that occur on the side of the star facing away from Earth. 
Then, at inferior conjunction, the lag and phase function will be zero, providing the worst possible detection conditions. 
Fortunately, transit and radial velocity detection methods work best under these conditions, so this regime is already accessible with existing techniques. 
Regardless, we must initially assign a weight, $w_{exo\_phase}$ of $\sim 50\%$ that there will be an echo, unless there are predictive factors like the position of active regions.
To account for phase-dependent exoplanet visibility, one weight choice is the Lambertian phase function (Eq. \ref{eq:lambertian}), $w_{exo\_phase}=\phi(\alpha)$.
In the case where an active region can be identified, a star-phase weight can be produced.
In preliminary simulations, we found that using a limb-darkening model appears to be effective, though considerable work remains to make these models effective.

Typically, all of these weights will be combined into a composite weight, e.g., 
\begin{equation}
W_{i}=w_{i, freq} w_{i, mag} w_{i, \Delta \omega} w_{i, exo\_phase}
\end{equation}
or, for the maximum likelihood approach, they may be combined as:
\begin{equation}
\sigma(W_{i})^2=w_{i, freq}^2 + w_{i, mag}^2 + w_{i, \Delta \omega}^2 + w_{i, exo\_phase}^2
\end{equation}
where the individual weights may require rescaling to have the appropriate relative balance.

\bibliographystyle{apj}
\bibliography{ms}{}

\end{document}